\documentclass[12pt,a4paper,fleqn]{article}
\usepackage[textheight=23cm,textwidth=16cm]{geometry}
\usepackage{amsmath}
\usepackage{graphicx}
\usepackage{cite}
\usepackage{hyperref}

\usepackage[american]{babel}

\usepackage{dcolumn} 
\newcolumntype{d}{D{.}{.}{2}}
\newcolumntype{e}{D{.}{.}{3}}
\newcolumntype{f}{D{.}{.}{4}}
\newcolumntype{g}{D{.}{.}{8}}

\usepackage{endfloat} 
\AtBeginDelayedFloats{\linespread{1.66}}

\newcommand{\be}{\boldsymbol{\mathcal{E}}}
\newcommand{\trel}[1]{\big\langle 0 \big| #1 \big| n \big\rangle}

\begin{document}

\begin{center}

{\LARGE\bf
  Origin-independent calculation of quadrupole intensities in X-ray spectroscopy \\
}

\vspace{2cm}

{\large 
Stephan Bernadotte, Andrew J. Atkins, and
Christoph R. Jacob\footnote{E-Mail: christoph.jacob@kit.edu},
}\\[4ex]

Karlsruhe Institute of Technology (KIT), \\ 
Center for Functional Nanostructures and Institute of Physical Chemistry, \\
Wolfgang-Gaede-Str. 1a, 76131 Karlsruhe, Germany

\vspace{9cm}

\vfil

\end{center}

\begin{tabbing}
Date:   \qquad\qquad\quad \= November 27, 2012 \\
Status:  \> published in \textit{J. Chem. Phys.} \textbf{137}, 204106 (2012).\\
DOI: \> \url{http://dx.doi.org/10.1063/1.4766359}
\end{tabbing}

\newpage

\begin{abstract}
  
For electronic excitations in the ultraviolet and visible range of the electromagnetic spectrum, the intensities
are usually calculated within the dipole approximation, which assumes that the oscillating electric field is
constant over the length scale of the transition. For the short wavelengths used in hard X-ray spectroscopy, the dipole
approximation may not be adequate. In particular, for metal K-edge X-ray absorption spectroscopy (XAS), it becomes 
necessary to include higher-order contributions. 
In quantum-chemical approaches to X-ray spectroscopy, these so-called quadrupole intensities have so far been calculated 
by including contributions depending on the square of the electric-quadrupole and magnetic-dipole transition moments.
However, the resulting quadrupole intensities depend on the choice of the origin of the coordinate system. Here,
we show that for obtaining an origin-independent theory, one has to include all contributions that are of the same 
order in the wave vector consistently. This leads to two additional contributions depending on products of the 
electric-dipole andelectric-octupole and  of the electric-dipole and magnetic-quadrupole transition moments, respectively. 
We have implemented such an origin-independent calculation of quadrupole intensities in XAS 
within time-dependent density-functional theory, and demonstrate its usefulness for the calculation of metal and 
ligand K-edge XAS spectra of transition metal complexes.

\end{abstract}

\newpage

\section{Introduction}

X-ray spectroscopy \cite{koningsberger_x-ray_1988,sthr_nexafs_1992,singh_advanced_2010} is a powerful 
spectroscopic tool for the elucidation of structural and electronic properties of materials \cite{guo_electronic_2011,
schuber_local_2011,gross_exafs_2010} and (bio-)molecular systems\cite{bauer_x-ray_2010,solomon_electronic_2004,
pollock_valence-to-core_2011}. X-ray absorption spectroscopy (XAS) probes the excitation of core electrons. Here, 
one distinguishes excitations to low-lying unoccupied states (so-called prepeaks), excitations to states close to the
ionization threshold (X-ray absorption near-edge structure, XANES), and excitations to continuum states
(extended x-ray absorption fine structure, EXAFS). On the other hand, X-ray emission spectroscopy (XES) 
considers the emission of X-ray radiation after the formation of a core hole. 

Of particular interest are the applications of these techniques to study catalytic mechanisms \textit{in situ} 
(for examples, see, Ref.~\cite{bauer_x-ray_2010,singh_advanced_2010,singh_electronic_2010,
kleymenov_structure_2012}) and to investigate biological or biomimetic systems containing transition metal 
centers (see, e.g., Refs.~\cite{ray_joint_2007,berry_electronic_2008,chandrasekaran_prediction_2011,sun_s_2011,
lancaster_x-ray_2011,beckwith_manganese_2011,roemelt_manganese_2012}). Usually, XAS and
XES are used as fingerprint techniques in such studies to identify the oxidation state, spin state, 
and local coordination environment of a metal center. This requires either the comparison with spectra 
measured for model compounds or with theoretical predictions. To be able to extract additional
information from X-ray spectroscopic measurements, the development of theoretical methods for
the calculation of X-ray spectra is essential \cite{carravetta_computational_2011}.

For EXAFS spectra, approaches based on scattering theory are well established and make it possible to 
extract structural parameters such as distances and coordination numbers \cite{rehr_theoretical_2000,
rehr_progress_2005}. In contrast, for describing prepeaks and the XANES region in XAS spectroscopy and 
for predicting XES spectra, quantum-chemical approaches are usually required. To this end, a wide range of 
quantum-chemical methods have been developed for describing excitations from core orbitals (for reviews, see, 
e.g., Refs~\cite{agren-rixs,carravetta_computational_2011,besley_time-dependent_2010}). Widely used are the 
static-exchange approximation (STEX) \cite{gren_direct_1994,gren_direct_1997,ekstrm_relativistic_2006}, approaches 
based on transition potential density-functional theory (DFT) \cite{triguero_calculations_1998,leetmaa_theoretical_2010}, 
and time-dependent density-functional theory (TD-DFT) \cite{besley_time-dependent_2010}. Within TD-DFT,
core excitations are accessible by restricting the space of occupied--virtual orbital excitations (restricted-channel
approximation) \cite{stener_time_2003,ray_description_2007}, by selectively targeting excitations within a specific 
energy window \cite{tsuchimochi_application_2008,schmidt_assignment_2010,kovyrshin_state-selective_2010,
liang_energy-specific_2011}, by using a complex polarization propagator \cite{ekstrm_polarization_2006,
ekstrm_x-ray_2006}, or with real-time TD-DFT methods \cite{lopata_linear-response_2012}. Recently,
coupled-cluster response theory has also been extended to X-ray spectroscopy \cite{coriani_coupled-cluster_2012,
coriani_asymmetric_2012}.

By combining X-ray spectroscopy with quantum-chemical calculations, it becomes possible to extract information
on the electronic structure of molecular systems. For instance, the analysis of the prepeak intensities in ligand 
K-edge XAS spectra of transition metal complexes (i.e., excitations from the ligand $1s$ to metal $d$ orbitals) 
provides insights into the covalent contributions to metal--ligand bonding \cite{shadle_ligand_1995,glaser_ligand_2000,nslund_direct_2003,
solomon_ligand_2005,debeer_george_metal_2005,debeer_george_calibration_2010,dey_s_2011}.
Another example is metal K-edge XAS, probing excitations from metal $1s$ to $d$ orbitals, which 
can be used to assign coordination numbers \cite{westre_multiplet_1997,debeer_george_prediction_2008,
hocking_fe_2009} and to probe details of the metal--ligand bonding mechanisms \cite{debeer_george_metal_2005,
scarborough_electronic_2011,roemelt_manganese_2012}. 
Recently, we have demonstrated that the prepeaks in Fe K-edge XAS spectra of ferrocene derivatives are 
sensitive to subtle differences in the electronic structure at the iron atom, which are induced by substituents
at the cyclopentadienyl rings, i.e., beyond the first coordination shell of the metal center \cite{atkins_probing_2012}.
Such studies are facilitated by high-energy resolution fluorescence detection (HERFD) techniques, which can
resolve the prepeaks with a much higher resolution than conventional XAS measurements \cite{hmlinen_elimination_1991,
de_groot_spectral_2002,glatzel_high_2005}.

However, beside the challenges posed by the accurate quantum-chemical prediction of the absolute or relative 
energies of core excitations, for the prepeaks in K-edge XAS spectra, theoretical X-ray spectroscopy faces an
additional problem. For the calculation of XAS intensities for hard X-rays, the well-known dipole approximation, in which the oscillator strengths are proportional to the square of the electric-dipole transition moments is not sufficient. 
The dipole
approximation is based on the assumption that the wavelength of the electromagnetic radiation is large compared
to the size of the core orbital. For the high-energy radiation used in hard X-ray spectroscopy, this is not the case
anymore. This is particularly important for the prepeaks in metal K-edge XAS spectra of transition metal complexes, 
which are usually dipole forbidden or have a very low dipole intensity due to symmetry. Thus, the intensity of these
prepeaks is due to contributions that are not included in the dipole approximation \cite{shulman_observations_1976,
drger_multipole_1988,yamamoto_assignment_2008}.

Currently, contributions to XAS or XES intensities beyond the dipole approximation, so-called quadrupole intensities, 
are calculated by including additional contributions that are proportional to the squares of the electric-quadrupole and
the magnetic-dipole transition moments \cite{debeer_george_time-dependent_2008,lee_probing_2010}. However,
these additional contributions depend on the choice of the origin of the coordinate system. This situation is not
satisfactory, as a physical observable should be origin independent. To rectify this, Neese and coworkers suggested
to choose the origin differently for each excitation such that these additional contributions are 
minimized \cite{debeer_george_time-dependent_2008}. Usually, this is equivalent to placing the origin on the atom 
at which the excitation takes place. However, in cases where the dipole intensities are very small compared to the
quadrupole intensities, this scheme sometimes places the origin far away from the excited core orbital \cite{atkins_probing_2012},
which seems unphysical and affects the resulting intensities significantly. Moreover, the scheme will also fail for
excitations from core orbitals that are delocalized over different atomic centers, a situation which occurs for ligand-edge
XAS spectra or for metal K-edge spectra in polynuclear transition metal clusters. 

Thus, a theoretical framework for the origin-independent calculation of quadrupole intensities in X-ray spectroscopy
would be desirable. Here, we show that such a formulation can be obtained if all contributions to the oscillator
strengths that are of the same order in the wave vector are included consistently.

This work is organized as follows. The theory is presented in Section~\ref{sec:theory}. After introducing the theoretical
framework in Sections~\ref{sec:radiation} and~\ref{sec:qm}, we revisit the multipole expansion of the transition
moments in Section~\ref{sec:multipole}. Subsequently, in Section~\ref{sec:osc-strengths} this expansion is applied 
for the calculation of the oscillator strengths and we demonstrate that these become origin-independent if all terms
that are of the same order are included consistently. The final equations for the isotropically averaged quadrupole
intensities are then derived in Section~\ref{sec:average}. This is followed by a description of our implementation
of the resulting formalism within TD-DFT in Section~\ref{sec:compdet}, before we illustrate its usefulness for two
test cases in Section~\ref{sec:results}. Finally a summary and concluding remarks are given in Sect.~\ref{sec:conclusion}

\section{Theory}
\label{sec:theory}

For the theoretical description of spectroscopic processes, quantum chemistry commonly employs a
semi-classical theory. In this framework, the molecules are described with (nonrelativistic) quantum-mechanics, 
whereas the electromagnetic radiation is treated classically (for a discussion, see also Ref.~\cite{reiher_relativistic_2009}).  
This theoretical framework is also appropriate for absorption and emission processes in X-ray spectroscopy.
Here, we will focus on the case of absorption, but the results can be transferred to other types of experiments.

\subsection{Electromagnetic Radiation}
\label{sec:radiation}

Within the Coulomb gauge (i.e., if one chooses the vector potential such that $\boldsymbol{\nabla}\cdot\boldsymbol{A}=0$),
a monochromatic, linearly-polarized electromagnetic wave is defined by the scalar and vector 
potentials \cite{jackson_classical_1998,cohen-tannoudji-2,schatz_quantum_2002},
\begin{align}
  \phi(\boldsymbol{r},t) &= 0 \\
    \label{eq:em-wave-a}
  \boldsymbol{A}(\boldsymbol{r},t) &= - A_0 \, \be \cos(\boldsymbol{k}\cdot\boldsymbol{r} - \omega t),
\end{align}
where the wave vector $\boldsymbol{k}$ points in the direction of propagation and its magnitude is related to the 
wavelength by $\lambda = 2\pi / k$, where $k=|\boldsymbol{k}|$. The angular frequency $\omega$ is $\omega=2\pi\nu$
with the frequency $\nu$, and frequency and wavelength are related by $c=\lambda\nu=\omega/k$, 
where $c$ is the speed of light. Finally, the polarization vector $\be$ is a real unit vector that is perpendicular to 
the direction of propagation (i.e., $\be\cdot\boldsymbol{k} = 0$). 

From these scalar and vector potentials, one obtains for the electric and magnetic fields,
\begin{align}
  \label{eq:em-wave-e}
  \boldsymbol{E}(\boldsymbol{r},t) 
  &= -\boldsymbol{\nabla}\phi(\boldsymbol{r},t) - \frac{1}{c} \frac{\partial\boldsymbol{A}(\boldsymbol{r},t)}{\partial t} 
  =  A_0 k \, \be \sin(\boldsymbol{k}\cdot\boldsymbol{r} - \omega t) 
  \\
  \label{eq:em-wave-b}
  \boldsymbol{B}(\boldsymbol{r},t) 
  &= \boldsymbol{\nabla}\times\boldsymbol{A}(\boldsymbol{r},t) 
  = A_0 (\boldsymbol{k}\times\be) \sin(\boldsymbol{k}\cdot\boldsymbol{r} - \omega t).
\end{align}
Here and in the following, we are using the Gaussian system of units. The electric and magnetic fields
are perpendicular to each other and to the direction of propagation and are oscillating with angular 
frequency $\omega$ and the wavelength $\lambda$.  The amplitudes of the electric and 
magnetic fields are $E_0 = B_0 = A_0 k$.

The intensity $I(\omega)$ of the electromagnetic radiation is defined as the energy flux per area through a 
surface perpendicular to the propagation direction. It can be calculated from the Poynting vector \cite{jackson_classical_1998},
\begin{equation}
  \boldsymbol{S} = \frac{c}{4\pi} (\boldsymbol{E}\times\boldsymbol{B}),
\end{equation}
by taking the absolute value and averaging over one period of the oscillations,
\begin{equation}
  \label{eq:em-wave-int}
  I(\omega) = \int_0^{1/\nu} |{\boldsymbol{S}}| \, {\rm d}t = \frac{1}{8\pi} \frac{\omega^2}{c} A_0^2 = \frac{c}{8\pi} k^2 A_0^2. 
\end{equation}

\subsection{Molecules in an Electromagnetic Field}
\label{sec:qm}

In the absence of an external electromagnetic field, a molecular system within the Born--Oppenheimer
approximation is described by the nonrelativistic Hamiltonian
\begin{align}
\label{eq:mol-hamiltonian}
  \hat{H}_0 
  = \sum_{i=1}^N \frac{\hat{\boldsymbol{p}}_i^2}{2m_e} + V(\boldsymbol{r}_1,\dotsc,\boldsymbol{r}_N),
\end{align}
where the momentum operator is given by $\hat{p}=-{\rm i}\hbar\boldsymbol{\nabla}$, $m_e$ and $e$ are the 
mass and the charge of the electron, respectively, and the potential energy $V(\boldsymbol{r}_1,\dotsc,\boldsymbol{r}_N)$
contains the electron-nuclei attraction as well as the electron--electron repulsion. Here and in the following, the 
index $i$ is used to label the electrons.

An external vector potential can be included in this Hamiltonian via \cite{schwabl,cohen-tannoudji-2,schatz_quantum_2002},
\begin{align}
  \hat{H} 
  &= \sum_i \frac{1}{2m_e} \Big[ \hat{\boldsymbol{p}}_i - \frac{e}{c} \boldsymbol{A}(\boldsymbol{r}_i,t) \Big]^2 
       - \frac{ge}{2m_ec} \sum_i \boldsymbol{B}(\boldsymbol{r}_i,t) \cdot \hat{\boldsymbol{s}}_i
       + V(\boldsymbol{r}_1,\dotsc,\boldsymbol{r}_N)
  \nonumber \\
  &= \sum_i \bigg[ \frac{\hat{\boldsymbol{p}}_i^2}{2m_e}   
       - \frac{e}{m_ec} \boldsymbol{A}(\boldsymbol{r}_i,t) \cdot \hat{\boldsymbol{p}}_i 
       + \frac{e^2}{2m_ec^2} \boldsymbol{A}^2(\boldsymbol{r}_i,t) \bigg]
  \nonumber \\
  &\hspace{3cm}     - \frac{ge}{2m_ec} \sum_i \boldsymbol{B}(\boldsymbol{r}_i,t) \cdot \hat{\boldsymbol{s}}_i
       + V(\boldsymbol{r}_1,\dotsc,\boldsymbol{r}_N),
\end{align}
where $g$ is the electron $g$-factor.
In the second line we used that in the Coulomb gauge, $\boldsymbol{p}_i\cdot\boldsymbol{A} = \boldsymbol{A}\cdot\boldsymbol{p}_i$.
After neglecting the term that is quadratic in $\boldsymbol{A}$, which is justified for weak electromagnetic fields,
this can be expressed as
\begin{equation}
  \hat{H} = \hat{H}_0 + \hat{U}(t),
\end{equation}
where the time-dependent perturbation is given by
\begin{align}
  \hat{U}(t) 
  &=  - \frac{e}{m_ec}  \sum_i \boldsymbol{A}(\boldsymbol{r}_i,t) \cdot \hat{\boldsymbol{p}}_i 
         - \frac{ge}{2m_ec} \sum_i \boldsymbol{B}(\boldsymbol{r}_i,t) \cdot \hat{\boldsymbol{s}}_i
\nonumber \\
  &=  \frac{eA_0 }{m_ec}  \sum_i \Big[ \cos(\boldsymbol{k}\cdot\boldsymbol{r}_i - \omega t) (\be \cdot \hat{\boldsymbol{p}}_i) 
         - \frac{g}{2} \sin(\boldsymbol{k}\cdot\boldsymbol{r}_i - \omega t) \, (\boldsymbol{k}\times\be) \cdot \hat{\boldsymbol{s}}_i \Big].
\end{align}
Here, we inserted the vector potential and the magnetic field of an electromagnetic wave given in
Eqs.~\eqref{eq:em-wave-a} and~\eqref{eq:em-wave-b}. Using $\sin(x)=\dfrac{1}{2{\rm i}} [\exp({\rm i}x) - \exp(-{\rm i}x)]$,
this can be expressed in the form
\begin{equation}
  \hat{U}(t) = \hat{U} \exp(-{\rm i}\omega t) + \hat{U}^* \exp({\rm i}\omega t),
\end{equation}
with the time-independent perturbation operator,
\begin{equation}
\hat{U} =  \frac{eA_0 }{2m_ec}  \sum_i \Big[ \exp({\rm i}\boldsymbol{k}\cdot\boldsymbol{r}_i) (\be \cdot \hat{\boldsymbol{p}}_i) 
         +{\rm i} \, \frac{g}{2} \exp({\rm i}\boldsymbol{k}\cdot\boldsymbol{r}_i) \, (\boldsymbol{k}\times\be) \cdot \hat{\boldsymbol{s}}_i 
         \Big].
\end{equation}

With this form of the perturbation, we can apply Fermi's golden rule to obtain the transition 
rate (i.e., the rate of change in the probability of finding the molecule in the $n$-th excited 
state) \cite{cohen-tannoudji-2,schatz_quantum_2002,atkins_molecular_2010}
\begin{equation}
  \Gamma_{0n}(\omega) 
  = \frac{2\pi}{\hbar} \big|\trel{\hat{U}}\big|^2 \ \delta(\omega-\omega_{0n})
  = \frac{\pi A_0^2}{2 \hbar c^2} \, |T_{0n}|^2 \ \delta(\omega-\omega_{0n}),
\end{equation}
where we introduced the transition moments
\begin{equation}
  \label{eq:t-full}
 T_{0n} = \frac{e}{m_e} \sum_i
  \trel{\exp(\text{i}\boldsymbol{k}\cdot\boldsymbol{r}_i)\,(\hat{\boldsymbol{p}}_i\cdot \be)
+\text{i} \, \frac{g}{2}  \, \exp(\text{i}\boldsymbol{k}\cdot\boldsymbol{r}_i)
   \, (\boldsymbol{k}\times \be) \cdot \hat{\boldsymbol{s}}_i \, }.
\end{equation}
Here, $|0\rangle$ and $|n\rangle$ are the eigenfunctions of the time-independent Hamiltonian 
$\hat{H}_0$ with $\hat{H}_0 |n\rangle = E_n |n\rangle$, and transitions only occur if the frequency 
of the perturbation matches the energy differences between eigenstates of the unperturbed 
molecule, i.e., for $\omega = \omega_{0n} = (E_n - E_0)/\hbar$.

Now, Eq.~\eqref{eq:em-wave-int} can be used to eliminate $A_0^2$ from the equation 
for the transition rate to arrive at,
\begin{equation}
  \Gamma_{0n}(\omega) 
  = \frac{4\pi^2}{c\hbar \omega^2} \, |T_{0n}|^2 I(\omega) \ \delta(\omega-\omega_{0n}).
\end{equation}
The absorption cross section, describing the rate of energy transfer from the electromagnetic
radiation to the molecule, is defined as
\begin{equation}
  \sigma_{0n} = \int \frac{\Gamma_{0n}(\omega) \hbar\omega}{I(\omega)} \, {\rm d}\omega = \frac{4\pi^2\hbar}{cE_{0n}} \, |T_{0n}|^2,
\end{equation}
where $E_{0n}=E_n-E_0$.
Finally, one usually introduces the dimensionless oscillator strengths,
\begin{equation}
  \label{eq:osci}
  f_{0n} = \frac{m_ec}{2\pi^2 e^2 \hbar} \sigma_{0n}
  =  \frac{2m_e}{e^2E_{0n}} \, |T_{0n}|^2. 
\end{equation}
These are defined as transition rates relative to a harmonic oscillator model \cite{schatz_quantum_2002,
atkins_molecular_2010}, which fixes the prefactor connecting the absorption cross section and the oscillator 
strengths.

\subsection{Multipole Expansion}
\label{sec:multipole}

Calculating the oscillator strengths via the matrix elements of Eq.~\eqref{eq:t-full} would in principle
be possible, but is cumbersome and in general not feasible. The required integrals are difficult to compute 
analytically (for a possible approach, see Ref.~\cite{lehtola_erkalea_2012}), 
and because of its dependence on the wave vector $\boldsymbol{k}$, the operator in $T_{0n}$ is different for each 
excitation. Therefore, one usually performs a multipole expansion. The starting point for this expansion
is a development of the exponential in a Taylor series,
\begin{equation}
  \exp(\text{i}\boldsymbol{k}\cdot\boldsymbol{r}_i)  = 1 + {\rm i} (\boldsymbol{k}\cdot\boldsymbol{r}_i) - \frac{1}{2}(\boldsymbol{k}\cdot\boldsymbol{r}_i)^2 + \dotsb
\end{equation}
This is substituted into Eq.~\eqref{eq:t-full} and, subsequently, one collects the terms of different orders in the 
wave vector $\boldsymbol{k}$, i.e., 
\begin{equation}
  T_{0n} = T_{0n}^{(0)} + T_{0n}^{(1)} + T_{0n}^{(2)} + \dotsb 
\end{equation}
In the following, we will consider terms up to second order in $\boldsymbol{k}$. Here, $|\boldsymbol{k}| = {2\pi}/{\lambda}$
acts as the expansion parameter, and we note that for larger wavelengths $\lambda$, the convergence of the Taylor 
expansion will be faster. For typical molecules and wavelengths in the ultraviolet or visible range, the wavelength
is large compared to the molecular size, and it is sufficient to include only the first (zeroth-order) term in this expansion. 
This corresponds to assuming that the oscillating electric field is constant over the whole molecule. However, for the 
short wavelengths used in hard X-ray spectroscopy this approximation is not adequate and higher-order terms need to be included.

\subsubsection{Zeroth order: Electric-dipole moment}

In zeroth order in the wave vector $\boldsymbol{k}$, we have
\begin{equation}
\label{eq:t-0th-order}
T^{(0)}_{0n} = \frac{e}{m_e}\sum_i \trel{\hat{\boldsymbol{p}}_i \cdot \be}
    = \be \cdot \trel{ \hat{\boldsymbol{\mu}}^p},
\end{equation}
where we have introduced the electric-dipole moment operator in the velocity representation
\begin{equation}
  \hat{\boldsymbol{\mu}}^p = \frac{e}{m_e} \sum_i \hat{\boldsymbol{p}}_i
\end{equation}

By using the relations given in Appendix \ref{app:comm}, the matrix elements of the electric-dipole moment 
operator in the velocity representation can be related to those in the conventional length representation as
\begin{equation}
 \trel{\hat{\boldsymbol{\mu}}^p} = - \frac{\rm i}{\hbar} E_{0n} \, \trel{\hat{\boldsymbol{\mu}} }, 
\end{equation}
where we introduced the electric-dipole moment operator in the length representation
\begin{equation}
  \hat{\boldsymbol{\mu}} = e \sum_i \hat{\boldsymbol{r}}_i.
\end{equation}

Thus, for the zeroth-order contribution, we arrive at
\begin{align}
T^{(0)}_{0n} = T^{(\mu)}_{0n} 
     = - \text{i} \frac{E_{0n}}{\hbar}  \, \Big( \be \cdot   \trel{\hat{\boldsymbol{\mu}} } \Big).
\end{align}

\subsubsection{First order: Electric-quadrupole and magnetic-dipole moments}

In the first order in $\boldsymbol{k}$, we find
\begin{equation}
\label{eq:t-1st-order}
T_{0n}^{(1)} =  \frac{{\rm i} \, e}{m_e}\sum_i \trel{ (\boldsymbol{k}\cdot\boldsymbol{r}_i) (\hat{\boldsymbol{p}}_i\cdot\be) }
                         + \frac{{\rm i} \, e g}{2 m_e} \sum_i \trel{ (\boldsymbol{k}\times\be) \cdot \hat{\boldsymbol{s}}_i }.
\end{equation}
The matrix elements in the first term can be split into one term that is symmetric and one that is antisymmetric with 
respect to interchanging the wave vector $\boldsymbol{k}$ and the polarization vector $\boldsymbol{\mathcal{E}}$ 
via
\begin{align}
 \trel{ (\boldsymbol{k}\cdot\boldsymbol{r}_i) (\hat{\boldsymbol{p}}_i \cdot \be) }
 =& \frac{1}{2} \trel{ (\boldsymbol{k}\cdot\boldsymbol{r}_i) (\hat{\boldsymbol{p}}_i\cdot\be) 
                                                                        + (\boldsymbol{k} \cdot \hat{\boldsymbol{p}}_i) (\boldsymbol{r}_i\cdot \be) }  \nonumber \\
 &+ \frac{1}{2} \trel{ (\boldsymbol{k}\cdot\boldsymbol{r}_i) (\hat{\boldsymbol{p}}_i \cdot \be) 
                                                                        - (\boldsymbol{k} \cdot \hat{\boldsymbol{p}}_i) (\boldsymbol{r}_i\cdot \be) }. 
\end{align}

From the symmetric first term and using Einstein's convention of implicit summation over 
repeated Greek indices, which we use to label the Cartesian components $x$, $y$, and $z$, we obtain
\begin{align}
  T_{0n}^{(Q)} =& \ \frac{{\rm i} \, e}{2m_e}\sum_i 
                                 k_\alpha \mathcal{E}_\beta  \trel{ r_{i,\alpha} \hat{p}_{i,\beta} + \hat{p}_{i,\alpha} r_{i,\beta} }  
                         =  \frac{{\rm i}}{2} \, k_\alpha \mathcal{E}_\beta \, \trel{ \hat{Q}^p_{\alpha\beta} }                          
\end{align}
where we introduced the electric-quadrupole moment operator in the velocity representation
\begin{equation}
  \hat{Q}^p_{\alpha\beta}  = \frac{e}{m_e} \sum_i \big(  r_{i,\alpha} \hat{p}_{i,\beta}  + \hat{p}_{i,\alpha} r_{i,\beta} \big).
\end{equation}

Again, with the help of the relations given in Appendix \ref{app:comm}, the matrix elements in the velocity representation 
can be related to those in the conventional length representation, and one arrives at
\begin{equation}
  T_{0n}^{(Q)} =  \frac{E_{0n}}{2 \hbar} \, k_\alpha \mathcal{E}_\beta  \, \trel{ \hat{Q}_{\alpha\beta} },                         
\end{equation}
where
\begin{equation}
  \label{eq:op-quad-length-def}
  \hat{Q}_{\alpha\beta}  = e \sum_i r_{i,\alpha}  r_{i,\beta} 
\end{equation}
is the operator of the electric-quadrupole moment in the length representation. Note that, in contrast to most other
authors \cite{barron-book,debeer_george_time-dependent_2008}, we do not introduce a traceless version of the 
quadrupole tensor here. The traceless definition arises from the expansion of $1/|\boldsymbol{r}|$ that is often 
introduced in the context of intermolecular interactions,  whereas in the case of an expansion of the exponential 
$\exp({\rm i}\boldsymbol{k}\cdot\boldsymbol{r})$ considered here the definition of Eq.~\eqref{eq:op-quad-length-def} is 
more natural. Nevertheless, because the wave vector $\boldsymbol{k}$ and the polarization vector $\be$ are orthogonal,
the diagonal elements of the electric-quadrupole transition moments do not enter here and it would, therefore, be 
possible to alter their trace without consequences.

For the antisymmetric second term, we can use that $(\boldsymbol{k}\cdot\boldsymbol{r}_i)$ and $(\hat{\boldsymbol{p}}_i \cdot \be)$ 
commute because $\boldsymbol{k}$ and $\be$ are orthogonal and then apply the vector identity 
\begin{equation}
 \label{eq:vectid}
(\boldsymbol{a}\cdot\boldsymbol{c})(\boldsymbol{b}\cdot\boldsymbol{d})  - (\boldsymbol{b}\cdot\boldsymbol{c})(\boldsymbol{a}\cdot\boldsymbol{d}) 
= (\boldsymbol{a}\times\boldsymbol{b})(\boldsymbol{c}\times\boldsymbol{d}) 
\end{equation}
to obtain,
\begin{align}
 T_{0n}^{(m')}  
 &= \frac{{\rm i} \, e}{2m_e}\sum_i \trel{ (\boldsymbol{k}\cdot\boldsymbol{r}_i) (\hat{\boldsymbol{p}}_i \cdot \be) 
                                                                        - (\boldsymbol{k} \cdot \hat{\boldsymbol{p}}_i) (\boldsymbol{r}_i\cdot \be) } 
 \nonumber \\                                                                       
 &= {\rm i} \frac{e}{2m_e}\sum_i \trel{ (\boldsymbol{k} \times \be)(\boldsymbol{r}_i \times \hat{\boldsymbol{p}}_i)}
 = {\rm i} c \, (\boldsymbol{k} \times \be) \cdot \trel{ \hat{\boldsymbol{m}}' },
\end{align}
with the (spin-independent) orbital magnetic-dipole moment operator
\begin{equation}
  \hat{\boldsymbol{m}}' = \frac{e}{2m_ec} \sum_i (\boldsymbol{r}_i \times \hat{\boldsymbol{p}}_i).
\end{equation}
Thus, this antisymmetric term $T_{0n}^{(m')}$ adopts the same form as the last, spin-dependent term in 
Eq.~\eqref{eq:t-1st-order},
\begin{equation}
  T_{0n}^{(m^s)} 
  = {\rm i} \frac{e}{2m_e}\sum_i (\boldsymbol{k} \times \be) \cdot \trel{g \, \hat{\boldsymbol{s}}_i } 
  = {\rm i} c \, (\boldsymbol{k} \times \be) \cdot \trel{ \hat{\boldsymbol{m}}^s},
\end{equation}
with the spin magnetic-dipole operator
\begin{equation}
  \hat{\boldsymbol{m}}^s = \frac{e}{2m_ec} \sum_i  g \, \hat{\boldsymbol{s}}_i.
\end{equation}
Combining the two contributions, we arrive at the magnetic-dipole transition moment,
\begin{equation}
  T_{0n}^{(m)} 
  = {\rm i} \frac{e}{2m_e}\sum_i (\boldsymbol{k} \times \be) \cdot \trel{(\boldsymbol{r}_i \times \hat{\boldsymbol{p}}_i) + g \, \hat{\boldsymbol{s}}_i }
  = {\rm i} c \, (\boldsymbol{k} \times \be) \cdot \trel{ \hat{\boldsymbol{m}} },
\end{equation}
and the total magnetic-dipole moment operator,
\begin{equation}
  \hat{\boldsymbol{m}} = \frac{e}{2m_ec} \sum_i  \bigg[ (\boldsymbol{r}_i \times \hat{\boldsymbol{p}}_i) + g \, \hat{\boldsymbol{s}}_i \bigg].
\end{equation}
Altogether, the first-order transition moments consist of an electric-quadrupole and a magnetic-dipole contribution, i.e.,
\begin{equation}
  T_{0n}^{(1)} = T_{0n}^{(Q)} + T_{0n}^{(m')} + T_{0n}^{(m^s)} = T_{0n}^{(Q)} + T_{0n}^{(m)}.
\end{equation}

\subsubsection{Second order: Electric-octupole and magnetic-quadrupole moments}

In second order in $\boldsymbol{k}$, we find,
\begin{equation}
\label{eq:t-2nd-order}
T_{0n}^{(2)} = - \frac{e}{2m_e}\sum_i \trel{ (\boldsymbol{k}\cdot\boldsymbol{r}_i)(\boldsymbol{k}\cdot\boldsymbol{r}_i)(\hat{\boldsymbol{p}}_i\cdot\be)}
- \frac{eg}{2m_e} \sum_i \trel{ (\boldsymbol{k}\cdot\boldsymbol{r}_i)(\boldsymbol{k}\times\be) \cdot \hat{\boldsymbol{s}}_i }.
\end{equation}
In a similar fashion as for the first-order term above, the matrix elements in the first term are split into a part that is symmetric 
and one that is antisymmetric with respect to interchanging the polarization vector $\be$ with one of the wave vectors 
$\boldsymbol{k}$,
\begin{align}
  \label{eq:second-order-matrixelements}
  &\trel{ (\boldsymbol{k}\cdot\boldsymbol{r}_i)(\boldsymbol{k}\cdot\boldsymbol{r}_i)(\hat{\boldsymbol{p}}_i\cdot\be)} \nonumber \\
  &\qquad= \frac{1}{3} \, \trel{ (\boldsymbol{k}\cdot\boldsymbol{r}_i)(\boldsymbol{k}\cdot\boldsymbol{r}_i)(\hat{\boldsymbol{p}}_i\cdot\be)
  \nonumber \\[-1ex]
  & \hspace{4cm}
                                                                    + (\boldsymbol{k}\cdot\boldsymbol{r}_i)(\boldsymbol{k}\cdot\hat{\boldsymbol{p}}_i)(\boldsymbol{r}_i\cdot\be)
                                                                    + (\boldsymbol{k}\cdot\hat{\boldsymbol{p}}_i)(\boldsymbol{k}\cdot\boldsymbol{r}_i)(\boldsymbol{r}_i\cdot\be)}
  \nonumber \\[1ex]                                                                  
  &\qquad +\frac{1}{3} \, \trel{ 2(\boldsymbol{k}\cdot\boldsymbol{r}_i)(\boldsymbol{k}\cdot\boldsymbol{r}_i)(\hat{\boldsymbol{p}}_i\cdot\be)
  \nonumber \\[-1ex]
  & \hspace{4cm}
                                                                    - (\boldsymbol{k}\cdot\boldsymbol{r}_i)(\boldsymbol{k}\cdot\hat{\boldsymbol{p}}_i)(\boldsymbol{r}_i\cdot\be)
                                                                    - (\boldsymbol{k}\cdot\hat{\boldsymbol{p}}_i)(\boldsymbol{k}\cdot\boldsymbol{r}_i)(\boldsymbol{r}_i\cdot\be)}.
\end{align}

For the symmetric first term, we obtain
\begin{align}
  T_{0n}^{(O)} &= - \frac{e}{6m_e}\sum_i k_\alpha k_\beta \mathcal{E}_\gamma \,
                             \trel{ r_{i,\alpha} r_{i,\beta} \hat{p}_{i,\gamma} + r_{i,\alpha} \hat{p}_{i,\beta} r_{i,\gamma} + \hat{p}_{i,\alpha} r_{i,\beta} r_{i,\gamma}}
  \nonumber \\
  &= - \frac{1}{6} \sum_i k_\alpha k_\beta \mathcal{E}_\gamma \trel{ \hat{O}^p_{\alpha\beta\gamma}},
\end{align}
with the operator of the electric-octupole moment in velocity representation
\begin{equation}
   \hat{O}^p_{\alpha\beta\gamma} = \frac{e}{m_e} \sum_i \big( r_{i,\alpha} r_{i,\beta} \hat{p}_{i,\gamma} 
                           + r_{i,\alpha} \hat{p}_{i,\beta} r_{i,\gamma} + \hat{p}_{i,\alpha} r_{i,\beta} r_{i,\gamma}\big).
\end{equation}
Using the relations given in Appendix \ref{app:comm}, these matrix elements in the velocity 
representation can be converted to those in the conventional length representation, 
\begin{equation}
 T_{0n}^{(O)} = {\rm i}  \frac{E_{0n}}{6\hbar} k_\alpha k_\beta \mathcal{E}_\gamma \trel{ \hat{O}_{\alpha\beta\gamma}},
\end{equation}
with the octupole operator in length representation given by
\begin{equation}
   \hat{O}_{\alpha\beta\gamma} = e \sum_i r_{i,\alpha} r_{i,\beta} r_{i,\gamma}.
\end{equation}
Again, note that our definition differs from the one given elsewhere \cite{barron-book}, as we do not introduce a traceless
 form here. 
In fact, for the Cartesian expansion of the exponential \mbox{$\exp({\rm i}\boldsymbol{k}\cdot\boldsymbol{r})$} it turns out that introducing 
such a traceless definition here is not possible because only the terms depending on the trace of the octupole moments, 
$\hat{O}_{\alpha\alpha\beta}$ will contribute to the isotropically-averaged oscillator strengths later on. In contrast, when describing 
intermolecular interactions starting from an expansion of $1/|\boldsymbol{r}|$, these terms are zero and do not appear.

After some algebra (see Appendix \ref{app:antisymm-2nd-order}), the antisymmetric part of Eq.~\eqref{eq:t-2nd-order} can 
be expressed as 
\begin{align}
  T_{0n}^{(\mathcal{M}')} &= - \frac{e}{6m_e} \sum_i (\boldsymbol{k} \times \be) \cdot \trel{ (\boldsymbol{k}\cdot\boldsymbol{r}_i) \cdot (\boldsymbol{r}_i \times \hat{\boldsymbol{p}}_i) 
                                         + (\boldsymbol{r}_i \times \hat{\boldsymbol{p}}_i)(\boldsymbol{k}\cdot\boldsymbol{r}_i) }
  \nonumber \\
  &= - \frac{e}{6m_e} \sum_i (\boldsymbol{k} \times \be)_\alpha \, k_\beta \ \trel{ r_{i,\beta} \, (\boldsymbol{r}_i \times \hat{\boldsymbol{p}}_i)_\alpha 
                                         + (\boldsymbol{r}_i \times \hat{\boldsymbol{p}}_i)_\alpha \, r_{i,\beta} }                                           
  \nonumber \\
  &= - \frac{c}{2} \ (\boldsymbol{k} \times \be)_\alpha \, k_\beta \ \trel{ \hat{\mathcal{M}'}_{\alpha\beta} },                                           
\end{align}
with the (spin-independent) orbital magnetic-quadrupole operator \cite{barron-book,graham_light_1990},
\begin{equation}
\hat{\mathcal{M}}'_{\alpha\beta} 
= \frac{e}{2m_ec} \sum_i \frac{2}{3} \, \Big(r_{i,\beta}(\boldsymbol{r}_i\times\hat{\boldsymbol{p}}_i)_\alpha + (\boldsymbol{r}_i\times\hat{\boldsymbol{p}}_i)_\alpha r_{i,\beta}\Big).
\end{equation}
Note that this operator is not symmetric with respect to interchanging $\alpha$ and $\beta$. 

The remaining spin-dependent part of Eq.~\eqref{eq:t-2nd-order} is given by
\begin{align}
T_{0n}^{(\mathcal{M}^s)} 
  &= - \frac{eg}{2m_e} \sum_i (\boldsymbol{k}\times\be) \cdot \trel{ (\boldsymbol{k}\cdot\boldsymbol{r}_i) \cdot \hat{\boldsymbol{s}}_i } 
  \nonumber \\
  &= - \frac{c}{2} \sum_i (\boldsymbol{k}\times\be)_\alpha \, k_\beta \  \trel{ \hat{\mathcal{M}}^s_{\alpha\beta} }
\end{align}
with the spin contribution to the magnetic-quadrupole operator
\begin{equation}
\hat{\mathcal{M}}^s_{\alpha\beta} = \frac{e}{2m_ec} \sum_i  g \Big(  r_{i,\beta} \, \hat{s}_{i,\alpha} \Big).
\end{equation}
Finally, the full magnetic-quadrupole contribution is obtained by adding the orbital and spin contributions to obtain
\begin{equation}
  T_{0n}^{(\mathcal{M})} = T_{0n}^{(\mathcal{M}')} + T_{0n}^{(\mathcal{M}^s)} 
  = - \frac{c}{2} \ (\boldsymbol{k} \times \be)_\alpha \, k_\beta \ \trel{ \hat{\mathcal{M}'}_{\alpha\beta} + \hat{\mathcal{M}}^s_{\alpha\beta} }.    
\end{equation}
Altogether, the second-order transition moments consist of an electric-octupole and a magnetic-quadrupole contribution, i.e.,
\begin{equation}
  T_{0n}^{(2)} = T_{0n}^{(O)} + T_{0n}^{(\mathcal{M}')} + T_{0n}^{(\mathcal{M}^s)} = T_{0n}^{(O)} + T_{0n}^{(\mathcal{M})}.
\end{equation}

\subsubsection{Summary of the multipole transition moments}

In summary, the multipole expansion of the transition moments of Eq.~\eqref{eq:t-full} up to second order results in 
five different contributions,
\begin{equation}
  T_{0n} = T_{0n}^{(\mu)} + T_{0n}^{(Q)} + T_{0n}^{(m)} + T_{0n}^{(O)} + T_{0n}^{(\mathcal{M})} + \dotsb
\end{equation} 
The expressions for these contributions are summarized in Table~\ref{tab:trans-moments}. In zeroth order,
one encounters the well-known electric-dipole transition moments. Starting from first order, an electric and
a magnetic contribution appear. In first order, these are the electric-quadrupole and magnetic-dipole transition
moments, whereas in second order one has the electric-octupole and magnetic-quadrupole transition moments.

\begin{table}
  \caption{Overview of the different contributions appearing in the multipole expansion of the transition moments 
   of Eq.~\eqref{eq:t-full} up to second order.}
  \label{tab:trans-moments}
  \begin{center}
  \begin{tabular}{cll}
    \hline\hline
    order & \multicolumn{1}{c}{electric} & \multicolumn{1}{c}{magnetic} \\
    \hline \\[-2.5ex]
    0 & $T^{(\mu)}_{0n} = -\text{i} \dfrac{E_{0n}}{\hbar}  \displaystyle\sum_\alpha  \mathcal{E}_\alpha  \trel{ \hat{\mu}_\alpha } $ & \multicolumn{1}{c}{--} \\[3ex]
    1 & $T_{0n}^{(Q)} =  \dfrac{E_{0n}}{2 \hbar} \displaystyle\sum_{\alpha\beta}  k_\alpha \mathcal{E}_\beta  \, \trel{ \hat{Q}_{\alpha\beta} }$ 
        & $T_{0n}^{(m)} = {\rm i} c \displaystyle\sum_\alpha  (\boldsymbol{k} \times \be)_\alpha \, \trel{ \hat{m}_\alpha }$ \\[3ex]
    2 & $T_{0n}^{(O)} = {\rm i} \dfrac{E_{0n}}{6\hbar}  \displaystyle\sum_{\alpha\beta\gamma}  k_\alpha k_\beta \mathcal{E}_\gamma \, \trel{ \hat{O}_{\alpha\beta\gamma}}$ 
        & $T_{0n}^{(\mathcal{M})}  = - \dfrac{c}{2} \displaystyle\sum_{\alpha\beta}  (\boldsymbol{k} \times \be)_\alpha \, k_\beta \ \trel{ \hat{\mathcal{M}}_{\alpha\beta}}$ \\[1.5ex]
    \hline\hline 
  \end{tabular}
  \end{center}
\end{table}

If we restrict our considerations to a spin--orbit coupling free framework in the absence of static external magnetic fields, the wavefunction
can always be chosen as a real function. In this case, the electric transition integrals, $\trel{\hat{\mu}_\alpha}$, $\trel{\hat{Q}_{\alpha\beta}}$,
and $\trel{\hat{O}_{\alpha\beta\gamma}}$, as defined here are always real. Moreover, for the magnetic transition moments, the spin contributions 
can be neglected because for states with a multiplicity larger than zero (i.e., for $S > 0$) the different $M_S$-components of the multiplet will 
be degenerate \cite{mcweeny_spins_2004,jacob_spin_2012}
and the components with $+M_S$ and $-M_S$ provide spin contributions to the magnetic transition moments that cancel 
each other. Therefore, these spin contributions, $\trel{\hat{m}^s_\alpha}$ and $\trel{\hat{\mathcal{M}}^s_{\alpha\beta}}$, will not be considered
further in the following. The remaining orbital contribution to the magnetic transition integrals, $\trel{\hat{m}_\alpha}$ and $\trel{\hat{\mathcal{M}}_{\alpha\beta}}$, 
are then purely imaginary. As a consequence, we notice that the zeroth-order and second-order transition moments 
are purely imaginary, whereas the 
first-order transition moments are purely real. This holds both for the electric and for the magnetic contributions.

\subsubsection{Origin dependence of the multipole transition moments}
\label{sec:shift-equation}

While the electric-dipole transition moments are independent of the choice of the origin of the coordinate systems, 
the higher-order transition moments are origin dependent \cite{barron-book}. We give derivations of the expressions for 
the  change of the electric-quadrupole and electic-octupole transition moments and of the magnetic-dipole and magnetic-quadrupole 
transition moments upon shifting the origin from $\boldsymbol{O}$ to $\boldsymbol{O}+\boldsymbol{a}$ in the Supplementary
Material \cite{jcp-suppmat}. Here, we only present the final results as required for the following discussions.

For the electric-quadrupole transition moments, one has
\begin{equation}
  \trel{\hat{Q}_{\alpha\beta}(\boldsymbol{O}+\boldsymbol{a})} 
  =  \trel{\hat{Q}_{\alpha\beta}(\boldsymbol{O})} - a_\beta \trel{\hat{\mu}_\alpha} - a_\alpha \trel{\hat{\mu}_\beta},
\end{equation}
whereas for the electric-octupole transition moments, the corresponding expression becomes,
\begin{align}
  \trel{\hat{O}_{\alpha\beta\gamma}(\boldsymbol{O}+\boldsymbol{a})} 
  =& \ \trel{\hat{O}_{\alpha\beta\gamma}(\boldsymbol{O})} \nonumber \\
        &- a_\gamma \trel{\hat{Q}_{\alpha\beta}(\boldsymbol{O})} - a_\beta \trel{\hat{Q}_{\alpha\gamma}(\boldsymbol{O})} - a_\alpha \trel{\hat{Q}_{\beta\gamma}(\boldsymbol{O})}
\nonumber \\
        &+ a_\alpha a_\beta \trel{\hat{\mu}_\gamma} + a_\alpha a_\gamma \trel{\hat{\mu}_\beta} + a_\beta a_\gamma \trel{\hat{\mu}_\alpha}.
\end{align}
Thus, the electric-quadrupole and octupole moments are only origin independent if all lower-order electric
transition moments vanish. The above expressions hold both for the length and for the velocity representation.

For the magnetic-dipole transition moment, a shift of the origin of the coordinate systems results in,
\begin{align}
  \label{eq:origin-dep-mdip}
  \trel{m'_\alpha(\boldsymbol{O}+\boldsymbol{a})} 
  &= \trel{m'_\alpha(\boldsymbol{O})} - \varepsilon_{\alpha\beta\gamma} a_\beta  \frac{1}{2c}  \trel{ \hat{\mu}^p_{\gamma} } \nonumber \\
  &= \trel{m'_\alpha(\boldsymbol{O})} + \frac{{\rm i}}{2} \, \varepsilon_{\alpha\beta\gamma} a_\beta  \frac{E_{0n}}{\hbar c}  \trel{ \hat{\mu}_{\gamma}},
\end{align}
where $\varepsilon_{\alpha\beta\gamma}$ is the Levi-Civita tensor. Thus, a term depending on the 
electric-dipole moment in the velocity representation arises. This can be converted into the length representation 
using the relation from Appendix~\ref{app:comm}. The origin dependence of the magnetic-dipole 
transition moments vanishes if the electric-dipole transition moment is zero.

Finally, for the magnetic-quadrupole transition moment the origin dependence is given by \cite{graham_light_1990,
graham_multipole_2000}
\begin{align}
  \trel{\mathcal{M}'_{\alpha\beta}(\boldsymbol{O}+\boldsymbol{a})} 
  =& \ \trel{\mathcal{M}'_{\alpha\beta}(\boldsymbol{O})}  \nonumber \\
  &- \frac{1}{3c} \, \varepsilon_{\alpha\gamma\delta} a_\gamma \trel{\hat{Q}^p_{\beta\delta}(\boldsymbol{O})} 
    +  \frac{2}{3c} \varepsilon_{\alpha\gamma\delta} \, a_\beta a_\gamma  \trel{\hat{\mu}^p_{\delta}}
  \nonumber \\
  &+ \frac{2}{3} \delta_{\alpha\beta} \, \Big( \boldsymbol{a} \cdot \trel{\boldsymbol{m}'(\boldsymbol{O})} \Big)
  - 2 a_\beta \trel{ m'_\alpha(\boldsymbol{O}) }.
\end{align}
Again, the electric-dipole and electric-quadrupole transition moments in the velocity representation can 
be converted to the length representation to finally arrive at
\begin{align}
  \label{eq:origin-dep-mquad}
  \trel{\mathcal{M}'_{\alpha\beta}(\boldsymbol{O}+\boldsymbol{a})} 
  =& \ \trel{\mathcal{M}'_{\alpha\beta}(\boldsymbol{O})}  \nonumber \\
  &+ \frac{{\rm i}}{3} \frac{E_{0n}}{\hbar c} \, \varepsilon_{\alpha\gamma\delta} a_\gamma \trel{\hat{Q}_{\beta\delta}(\boldsymbol{O})} 
    - \frac{2{\rm i}}{3} \frac{E_{0n}}{\hbar c} \, \varepsilon_{\alpha\gamma\delta} \, a_\beta a_\gamma  \trel{\hat{\mu}_{\delta}}
  \nonumber \\
  &+ \frac{2}{3} \delta_{\alpha\beta} \, \Big( \boldsymbol{a} \cdot \trel{\boldsymbol{m}'(\boldsymbol{O})} \Big)
  - 2 a_\beta \trel{ m'_\alpha(\boldsymbol{O}) }.
\end{align}
Here, we notice that upon shifting the origin, the magnetic-quadrupole transition moment generates all lower-order
contributions, i.e., terms depending on the electric-dipole and electric-quadrupole transition moments as well as on
the magnetic-dipole transition moments.

\subsection{Oscillator Strengths}
\label{sec:osc-strengths}

The multipole expansion of the full transition moments $T_{0n}$ can now be inserted into Eq.~\eqref{eq:osci}
to obtain an expression for calculating the oscillator strengths,
\begin{align}
  \label{eq:osc-strengths-full}
   f_{0n} 
   &= \frac{2m_e}{e^2E_{0n}}\, \big|T_{0n}^{(0)} + T_{0n}^{(1)} + T_{0n}^{(2)} + \dotsb \big|^2 \nonumber \\
   &= \frac{2m_e}{e^2E_{0n}}\, \big|T_{0n}^{(\mu)} + T_{0n}^{(Q)} + T_{0n}^{(m)} + T_{0n}^{(O)} + T_{0n}^{(\mathcal{M})} + \dotsb \big|^2.
\end{align}
Here, different truncations of the expansion can be employed. In the \textit{dipole approximation}, only the zeroth-order
term is retained, and for the oscillator strengths one arrives at the well-known expression in which the oscillator 
strengths are proportional to the squared absolute value of the electric-dipole transition moments. This approximation
is commonly employed in electronic spectroscopy in the ultraviolet and visible region of the electromagnetic spectrum.

Here, we are interested in cases where the dipole approximation breaks down, such as K-edge XAS spectroscopy
of transition metal complexes. In such situations, higher-order terms in the multipole expansion have to be included. In the
currently used approximation, the multipole expansion of the transition moments is truncated after the first order, i.e., the
oscillator strengths are approximated as \cite{debeer_george_time-dependent_2008}
\begin{equation}
  \label{eq:osc-neese}
  f_{0n} \approx \frac{2m_e}{e^2E_{0n}}\, \big|T_{0n}^{(0)} + T_{0n}^{(1)} \big|^2.
\end{equation}   
However, it turns out that the resulting expressions depend on the choice of the origin of the coordinate system
(see Ref.~\cite{debeer_george_time-dependent_2008} for details).

\begin{figure}
\caption{Schematic illustration of the different terms arising from the squared absolute value in Eq.~\eqref{eq:osc-strengths-full}.
 The entries in the table indicate the order of the different terms in the wave vector $\boldsymbol{k}$. We retain all terms
 up to second order, as indicated by the red line.}
\label{fig:orders}
\begin{center}
\includegraphics{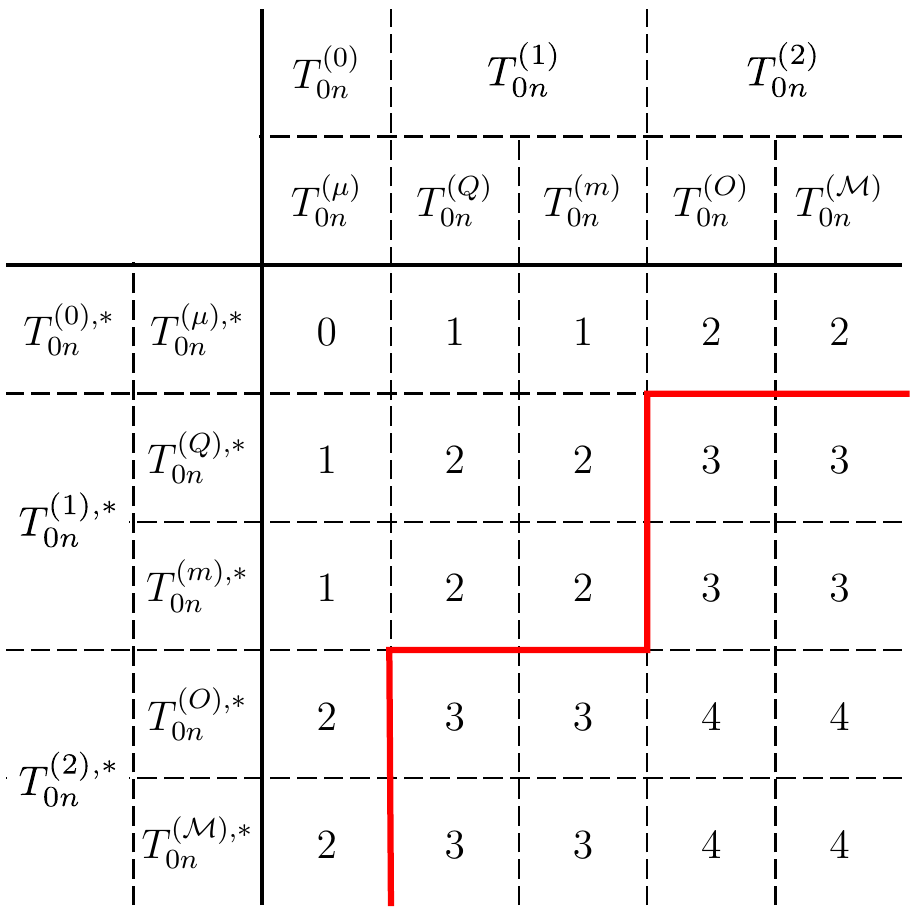}
\end{center}
\end{figure}

To obtain an origin-independent formulation, we return to Eq.~\eqref{eq:osc-strengths-full} and realize that the
squared absolute value results in a sum of products of multipole transition moments. These products are of different 
orders in the wave vector $\boldsymbol{k}$, as is illustrated in Fig.~\ref{fig:orders}. Hence, it seems logical to retain all 
terms up to a given order in $\boldsymbol{k}$ in the expression for the oscillator strengths instead of truncating the 
multipole expansion of the transition moments. By collecting terms that are of the same order, the oscillator strengths
can be expressed as
\begin{equation}
  f_{0n} = f_{0n}^{(0)} + f_{0n}^{(1)} + f_{0n}^{(2)} + \dotsb 
\end{equation}
where
\begin{align}
 f_{0n}^{(0)} &= \frac{2m_e}{e^2E_{0n}} \, \big|T_{0n}^{(0)}\big|^2 \\
 f_{0n}^{(1)} &= \frac{2m_e}{e^2E_{0n}} \, 2\text{Re}\big(T_{0n}^{(0),*} \, T_{0n}^{(1)}\big) = 0\\
 f_{0n}^{(2)} &= \frac{2m_e}{e^2E_{0n}} \, \Big[ \big|T_{0n}^{(1)}\big|^2 +  2\text{Re}\big( T_{0n}^{(0),*} \, T_{0n}^{(2)} \big) \Big],
\end{align}
where the star denotes complex conjugation. Because $T_{0n}^{(0)}$ is purely imaginary and $T_{0n}^{(1)}$
is real, their product is also purely imaginary and the first-order contribution $f_{0n}^{(1)}$ vanishes. In the following, 
we will retain all terms up to second order and it turns out that the resulting approximation for the oscillator strengths 
is independent of the choice of the origin.

\subsubsection{Origin independence of oscillator strengths}
\label{sec:osc-origin-indep}

Starting from the definitions of $T_{0n}^{(0)}$, $T_{0n}^{(1)}$, and $T_{0n}^{(2)}$ in Eqs.~\eqref{eq:t-0th-order},
\eqref{eq:t-1st-order}, and \eqref{eq:t-2nd-order}, respectively, one can easily see that their origin dependence is 
given by
\begin{align}
 T_{0n}^{(0)}(\boldsymbol{O} + \boldsymbol{a})  &= T_{0n}^{(0)}(\boldsymbol{O}), \\
  T_{0n}^{(1)}(\boldsymbol{O} + \boldsymbol{a}) 
  &=  T_{0n}^{(1)}(\boldsymbol{O}) +  {\rm i} (\boldsymbol{k}\cdot \boldsymbol{a}) \, T_{0n}^{(0)}(\boldsymbol{O}), \\
  T_{0n}^{(2)}(\boldsymbol{O} + \boldsymbol{a}) 
  &= T_{0n}^{(2)}(\boldsymbol{O}) + {\rm i} (\boldsymbol{k}\cdot \boldsymbol{a}) \, T_{0n}^{(1)}(\boldsymbol{O}) 
        - \frac{1}{2} (\boldsymbol{k}\cdot  \boldsymbol{a})^2 \, T_{0n}^{(0)}(\boldsymbol{O}). 
\end{align}
Therefore, the zeroth-order contribution to the oscillator strengths $f_{0n}^{(0)}$, i.e., the expression obtained 
in the dipole approximation, is obviously origin independent. For the second-order contribution, we have
\begin{align}
  f_{0n}^{(2)}(\boldsymbol{O} + \boldsymbol{a})  
  =& \ \frac{2m_e}{e^2E_{0n}}\, \bigg[ \big|T_{0n}^{(1)}(\boldsymbol{O} + \boldsymbol{a})\big|^2 
       +  2\text{Re}\Big( T_{0n}^{(0),*}(\boldsymbol{O}) \, T_{0n}^{(2)}(\boldsymbol{O} + \boldsymbol{a}) \Big) \bigg]\nonumber \\
  =& \ \frac{2m_e}{e^2E_{0n}}\, \bigg[ \big|T_{0n}^{(1)}(\boldsymbol{O})\big|^2 
       + 2\text{Re}\Big({\rm i} (\boldsymbol{k}\cdot \boldsymbol{a}) T_{0n}^{(0)}(\boldsymbol{O}) T_{0n}^{(1),*}(\boldsymbol{O})\Big) 
       + (\boldsymbol{k}\cdot \boldsymbol{a})^2 \, \big|T_{0n}^{(0)}(\boldsymbol{O})\big|^2
  \nonumber \\
  &\hspace{1.5cm}+  \text{Re} \Big( T_{0n}^{(0),*}(\boldsymbol{O}) \, \big[2T_{0n}^{(2)}(\boldsymbol{O}) 
     + 2{\rm i} (\boldsymbol{k}\cdot \boldsymbol{a}) \, T_{0n}^{(1)}(\boldsymbol{O}) 
     - (\boldsymbol{k}\cdot  \boldsymbol{a})^2 \, T_{0n}^{(0)}(\boldsymbol{O}) \big] \Big) \bigg]
  \nonumber \\
  =&  \ f_{0n}^{(2)}(\boldsymbol{O}) 
\end{align}
and find that this contribution is indeed independent of the choice of the origin. In fact, it can be shown that 
for each order, the higher-order contributions to the oscillator strengths are origin independent if all terms that are of the same 
order in the wave vector $\boldsymbol{k}$ are included. This is demonstrated in Appendix~\ref{app:origin-higher-order}.

\subsubsection{Dipole and quadrupole oscillator strengths}

After having established an origin-independent definition of the different approximations to the oscillator strengths, we
will now turn to deriving explicit expressions. Considering only the zeroth-order contribution corresponds to the dipole
approximation, in which the \textit{dipole oscillator strengths} are given by
\begin{equation}
  f_{0n} \approx f_{0n}^{(0)} = f_{0n}^{(\mu^2)}  
  = \frac{2m_e}{e^2\hbar^2} \,  E_{0n} \, \big| \be \cdot \trel{ \hat{\boldsymbol{\mu}} } \big|^2 
  = \frac{2m_e}{e^2\hbar^2} \,  E_{0n} \, \Big( \mathcal{E}_\alpha \trel{ \hat{\mu}_\alpha } \Big)^2.
\end{equation}

Since the first-order contributions vanish, the next step to go beyond the dipole approximation is to include all second-order 
contributions. Thus, the oscillator strengths can be approximated as the sum of the dipole (zeroth-order) oscillator strengths 
and the \textit{quadrupole (second-order) oscillator strengths}, 
\begin{equation}
  \label{eq:osc-full}
  f_{0n} \approx f_{0n}^{(0)} + f_{0n}^{(2)} =  \big|T_{0n}^{(0)}\big|^2 + \big|T_{0n}^{(1)}\big|^2 +  2\text{Re}\big(T_{0n}^{(0),*}T_{0n}^{(2)}\big).
\end{equation}
We will refer to this approximation as the \textit{quadrupole approximation}. For the quadrupole oscillator strengths, we can insert 
the individual multipole transition moments, and obtain five different terms,
\begin{align}
  f_{0n}^{(2)} &= \frac{2m_e}{e^2E_{0n}}\, \Big[ \big|T_{0n}^{(Q)}\big|^2 +  \big|T_{0n}^{(m)}\big|^2 
                           + 2 \text{Re}\big( T_{0n}^{(Q),*} T_{0n}^{(m)} \big) 
                          + 2 \text{Re}\big( T_{0n}^{(\mu),*} T_{0n}^{(O)} \big) + 2 \text{Re}\big( T_{0n}^{(\mu),*} T_{0n}^{(\mathcal{M})} \big) \Big] \nonumber \\  
 &=  f_{0n}^{(Q^2)} + f_{0n}^{(m^2)} + f_{0n}^{(Qm)} + f_{0n}^{(\mu O )} + f_{0n}^{(\mu \mathcal{M})}.
\end{align}
First, there are three contributions arising from products of first-order transition moments, an electric-quadrupole--electric-quadrupole
contribution,
\begin{align}
  f_{0n}^{(Q^2)} = \frac{m_e}{2 e^2\hbar^2} E_{0n} \, \Big( k_\alpha \mathcal{E}_\beta  \, \trel{ \hat{Q}_{\alpha\beta} } \Big)^2,
\end{align}
a magnetic-dipole--magnetic-dipole contribution,
\begin{align}
  f_{0n}^{(m^2)} = \frac{2m_ec^2 }{e^2E_{0n}} \, \Big| (\boldsymbol{k} \times \be)_\alpha \, \trel{ \hat{m}_\alpha } \Big|^2
                            = \frac{2m_ec^2 }{e^2E_{0n}} \, \Big( (\boldsymbol{k} \times \be)_\alpha \, \text{Im}\trel{ \hat{m}_\alpha } \Big)^2,
\end{align}
and a cross-term, the electric-quadrupole--magnetic-dipole contribution,
\begin{align}
  f_{0n}^{(Qm)} &= -\frac{2m_ec}{e^2 \hbar} \, \Big( k_\alpha \mathcal{E}_\beta  \, \trel{ \hat{Q}_{\alpha\beta} } \Big)
                                                                       \Big( (\boldsymbol{k} \times \be)_\alpha \, \text{Im}\trel{ \hat{m}_\alpha } \Big).
\end{align}
These three contributions have been considered previously in the calculation of the quadru\-pole oscillator strengths in 
Ref.~\cite{debeer_george_time-dependent_2008}.
In addition, two additional contributions have to be included in order to collect all terms that are of second order and
to arrive at an origin-independent approximation. These are the electric-dipole--electric-octupole contribution,
\begin{align}
  f_{0n}^{(\mu O)} &= - \frac{2m_e}{3 \hbar^2 e^2} E_{0n} \, \Big(\mathcal{E}_\alpha  \trel{ \hat{\mu}_\alpha } \Big)
                                     \Big(k_\alpha k_\beta \mathcal{E}_\gamma \, \trel{ \hat{O}_{\alpha\beta\gamma}}\Big),
\end{align}
and the electric-dipole--magnetic-quadrupole contribution,
\begin{align}
  f_{0n}^{(\mu \mathcal{M})} &= \frac{2m_ec}{e^2\hbar} \, \Big(\mathcal{E}_\alpha  \trel{ \hat{\mu}_\alpha } \Big)
                                     \Big( (\boldsymbol{k} \times \be)_\alpha \, k_\beta \ \text{Im}\trel{ \hat{\mathcal{M}}_{\alpha\beta}}\Big).
\end{align}

Now we choose the wave vector as $\boldsymbol{k}=k\boldsymbol{e}_x$ along the $x$-axis and the 
polarization vector as $\be = \boldsymbol{e}_y$ along the $y$-axis. Consequently, $(\boldsymbol{e}_x \times \be)$ 
becomes the unit vector $\boldsymbol{e}_z$ along the $z$-axis. This is no loss of generality, as the molecule 
can still have an arbitrary orientation in the coordinate system. Using 
\begin{equation}
  k = \frac{E_{0n}}{\hbar c},
\end{equation}
the different contributions to the oscillator strengths become
\begin{align}
   f_{0n}^{(\mu^2)}  
  &= \frac{2m_e}{e^2\hbar^2} \,  E_{0n} \, \trel{ \hat{\mu}_y }^2  
  \displaybreak[3] \\[1ex]
  f_{0n}^{(Q^2)} 
  &= \frac{m_e}{2 e^2\hbar^4 c^2} E_{0n}^3 \, \trel{ \hat{Q}_{xy} }^2
  \displaybreak[3] \\[1ex]
  f_{0n}^{(m^2)} 
  &= \frac{2m_e }{e^2\hbar^2} E_{0n} \, \Big[ \text{Im}\trel{ \hat{m}_z } \Big]^2
  \displaybreak[3] \\[1ex]
  f_{0n}^{(Qm)} 
  &= - \frac{2m_e}{e^2 \hbar^3 c} E_{0n}^2 \,  \trel{ \hat{Q}_{xy} } \text{Im}\trel{ \hat{m}_z } 
  \displaybreak[3] \\[1ex]
  f_{0n}^{(\mu O)} 
  &= - \frac{2m_e}{3  e^2 \hbar^4c^2} E_{0n}^3 \,  \trel{ \hat{\mu}_y } \trel{ \hat{O}_{xxy}}
  \displaybreak[3] \\[1ex]
  f_{0n}^{(\mu \mathcal{M})} 
  &= \frac{2m_e}{e^2\hbar^3 c} E_{0n}^2\, \trel{ \hat{\mu}_y } \text{Im}\trel{ \hat{\mathcal{M}}_{zx}}.
\end{align}
These oscillator strengths refer to an experimental setup in which the incident radiation has
a well-defined polarization and in which the molecules have a fixed orientation with respect
to the radiation. Using the expressions for the origin dependence of the different multipole
transition moments given in Section~\ref{sec:shift-equation}, it can be verified that the total 
second-order oscillator strengths calculated using the above equations are origin-independent 
(see Supplementary Material \cite{jcp-suppmat}).

\subsection{Isotropic Averaging}
\label{sec:average}

Often, the molecules are not oriented with respect to the incident radiation in experiments,
but the measurement is performed in solution where the molecules can freely rotate. Thus, to
arrive at final expressions for the oscillator strengths in such experiments, we have to perform
an averaging over all possible orientations of the molecule.  

The expressions for performing this averaging are derived, for instance, in Ref.~\cite{barron-book} (see in particular 
chapter 4.2). For the isotropic averages of tensors with two, three, and four Cartesian indices, one finds 
\begin{align}
  \langle T_{xx} \rangle_\text{iso} 
  &= \sum_{\alpha\beta} \langle i_\alpha i_\beta \rangle_\text{iso} \, T_{\alpha\beta} \\
  \langle T_{xyz} \rangle_\text{iso} 
  &= \sum_{\alpha\beta\gamma} \langle i_\alpha j_\beta k_\gamma \rangle_\text{iso} \, T_{\alpha\beta\gamma} \\
   \langle T_{xxyy} \rangle_\text{iso} 
   &= \sum_{\alpha\beta\gamma\delta} \langle i_\alpha i_\beta j_\gamma j_\delta \rangle_\text{iso} \, T_{\alpha\beta\gamma\delta},
\end{align}
where the isotropic averages of the Cartesian unit vector $\boldsymbol{i}=\boldsymbol{e}_x$, $\boldsymbol{j}=\boldsymbol{e}_y$, 
and $\boldsymbol{k}=\boldsymbol{e}_z$ are given by
\begin{align}
\langle i_\alpha i_\beta \rangle_\text{iso} 
&= \frac{1}{3}\delta_{\alpha\beta}\,, \\
\langle i_\alpha j_\beta k_\gamma \rangle_\text{iso} 
&= \frac{1}{6}\epsilon_{\alpha\beta\gamma}\,, \\
\langle i_\alpha i_\beta j_\gamma j_\delta \rangle_\text{iso} 
&= \frac{1}{30}\left(4\delta_{\alpha\beta}\delta_{\gamma\delta}
   -\delta_{\alpha\gamma}\delta_{\beta\delta}-\delta_{\alpha\delta}\delta_{\beta\gamma}\right).
\end{align}
For all other tensor components, such as, e.g., $ \langle T_{xy} \rangle_\text{iso}$ or $\langle T_{xxy} \rangle_\text{iso}$,
the isotropic averages are zero. 

Using these expressions, we obtain for the isotropically averaged electric-dipole--electric-dipole 
contribution to the oscillator strengths,
\begin{equation}
  \label{eq:osc-avg-dipole}
   \langle f_{0n}^{(\mu^2)} \rangle_\text{iso}  
  = \frac{2m_e}{3e^2\hbar^2} \,  E_{0n} \, \sum_\alpha \trel{ \hat{\mu}_\alpha }^2
  = \frac{2m_e}{3e^2\hbar^2} \, E_{0n} \, \trel{ \hat{\boldsymbol{\mu}} } ^2. 
\end{equation}

Similarly, for the electric-quadrupole--electric-quadrupole contribution, we find
\begin{align}
  \langle f_{0n}^{(Q^2)} \rangle_\text{iso}  &= \frac{m_e}{60 e^2\hbar^4 c^2} E_{0n}^3 \, 
                              \sum_{\alpha\beta\gamma\delta} \big( 4 \delta_{\alpha\gamma}\delta_{\beta\delta} 
                                          - \delta_{\alpha\beta}\delta_{\gamma\delta} - \delta_{\alpha\delta}\delta_{\beta\gamma} \big) 
                              \trel{ \hat{Q}_{\alpha\beta} } \trel{ \hat{Q}_{\gamma\delta} } \nonumber \\
  &= \frac{m_e}{20 e^2\hbar^4 c^2} E_{0n}^3 \, \bigg[  \sum_{\alpha\beta}  \trel{ \hat{Q}_{\alpha\beta} }^2 
                               - \frac{1}{3} \Big( \sum_{\alpha} \trel{ \hat{Q}_{\alpha\alpha} } \Big)^2 \bigg].                                                   
\end{align}
We note that this is identical to the expression in Ref.~\cite{debeer_george_time-dependent_2008}, where a traceless 
definition of the quadrupole moment is used. For the magnetic-dipole--magnetic-dipole contribution, the isotropic 
average is,
\begin{align}
  \langle f_{0n}^{(m^2)} \rangle_\text{iso}  
  = \frac{2m_e }{3e^2\hbar^2} E_{0n} \, \sum_\alpha \text{Im}\trel{ \hat{m}_\alpha }^2 
  = \frac{2m_e }{3e^2\hbar^2} E_{0n} \, \Big( \text{Im}\trel{ \hat{\boldsymbol{m}} } \Big)^2.
\end{align}
The isotropic average of the electric-quadrupole--magnetic-dipole contribution to the oscillator 
strengths,
\begin{align}
  \langle f_{0n}^{(Qm)} \rangle_\text{iso}
  &= -\frac{m_e}{3e^2 \hbar^3 c} E_{0n}^2 \,  \sum_{\alpha\beta\gamma} \varepsilon_{\alpha\beta\gamma} \trel{ \hat{Q}_{\alpha\beta} } \trel{ \hat{m}_\gamma }  = 0
\end{align}
turns out to be zero because $\trel{ \hat{Q}_{\alpha\beta} } = \trel{ \hat{Q}_{\beta\alpha} }$.
Finally, for the electric-dipole--electric-octupole contribution to the oscillator strengths, we obtain
\begin{align}
  \langle f_{0n}^{(\mu O)} \rangle_\text{iso}  &= - \frac{m_e}{45  e^2 \hbar^4c^2} E_{0n}^3 \, 
                                       \sum_{\alpha\beta\gamma\delta} \big( 4 \delta_{\alpha\beta}\delta_{\gamma\delta} 
                                                - \delta_{\alpha\gamma}\delta_{\beta\delta} - \delta_{\alpha\delta}\delta_{\beta\gamma} \big)  \trel{ \hat{\mu}_\delta } \trel{ \hat{O}_{\alpha\beta\gamma}} 
   \nonumber \\                                                
  &= - \frac{2m_e}{45  e^2 \hbar^4c^2} E_{0n}^3 \, \sum_{\alpha\beta} \trel{ \hat{\mu}_\beta } \trel{ \hat{O}_{\alpha\alpha\beta}}, 
\end{align}
where we used the symmetry of the octupole moments with respect to the exchange of indices, and for the
electric-dipole--magnetic-quadrupole contribution,
\begin{align}
    \label{eq:osc-avg-mquad}
  \langle f_{0n}^{(\mu \mathcal{M})} \rangle_\text{iso}  
 &= \frac{m_e}{3 e^2\hbar^3 c} E_{0n}^2\, 
         \sum_{\alpha\beta\gamma} \varepsilon_{\alpha\beta\gamma} \trel{ \hat{\mu}_\beta } \, \text{Im}\trel{ \hat{\mathcal{M}}_{\gamma\alpha}}. 
\end{align}
Note again that the magnetic-quadrupole transition moments are in general not symmetric or antisymmetric with respect to the 
interchange of the two Cartesian indices, i.e., $\trel{\mathcal{M}_{\alpha\beta}} \ne \pm \trel{\mathcal{M}_{\beta\alpha}}$.

In summary, there are five contributions to the isotropically averaged oscillator strengths up to second order.
Also at this stage it can be verified that the resulting total oscillator strengths are independent of the choice of
the origin, which is shown in the Supplementary Material \cite{jcp-suppmat}. Note that the individual contributions 
are still origin dependent. Therefore, a separation into electric and magnetic contributions will also depend on the 
choice of the origin.

\section{Computational Methodology and Implementation}
\label{sec:compdet}

The theory presented here for the origin-independent calculation of quadrupole oscillator
strengths is applicable in combination with any quantum-chemical method that is capable 
of providing excited states, either via a time-independent formulation or with response theory 
(for a review, see, e.g., Ref.~\cite{gomes_quantum-chemical_2012}). Here, we select TD-DFT 
which has become an important tool in computational X-ray spectroscopy in the past 
years \cite{besley_time-dependent_2009,besley_time-dependent_2010,
roemelt_manganese_2012,neese_prediction_2009}.

We have implemented the calculation of the second-order oscillator strengths into the TD-DFT 
module \cite{adf-tddft,autschbach_calculating_2002,autschbach_chiroptical_2002}
of the Amsterdam density functional (\textsc{Adf}) program package \cite{adf,chem-with-adf}.
Within TD-DFT, the required electric and magnetic transition moments are calculated as products 
of the solution vectors $(\boldsymbol{X}+\boldsymbol{Y})$ and $(\boldsymbol{X}-\boldsymbol{Y})$,
respectively, with the corresponding matrix elements in the basis of Kohn-Sham molecular orbitals
(for details, see, e.g., Refs.~\cite{tddft-3,furche_density_2001,autschbach_calculating_2002}). The 
required electric-octupole  and magnetic-quadrupole integrals are provided 
by \textsc{Adf}'s \textsc{AOResponse} module \cite{krykunov_calculation_2005,krykunov_calculation_2007}.

So far, we have only shown that the theory presented here is origin-independent for the exact eigenfunctions
of $\hat{H}_0$. However, an additional difficulty arises in approximate calculations. For deriving the equations
for the origin-dependence of the magnetic-dipole and magnetic-quadrupole transition moments [Eqs.~\eqref{eq:origin-dep-mdip}
and~\eqref{eq:origin-dep-mquad}], we have converted the occurring electric-dipole and electric-quadrupole transition moments
from the velocity to the length representation. However, this is only exact in the case of a complete, infinite basis set. Thus, in
calculations using a finite basis set, the magnetic-dipole and magnetic-quadrupole transition moments do not show the exact
origin dependence of Eqs.~\eqref{eq:origin-dep-mdip} and~\eqref{eq:origin-dep-mquad}.

Therefore, we calculate the electric-dipole, electric-quadrupole, and electric-octupole transition moments in the velocity representation for the 
second-order contributions to the oscillator strengths (see also Refs~\cite{escf-4,autschbach_calculating_2002} 
for the calculation of transition moments in the velocity representation with TD-DFT). It can be easily verified that 
this results in second-order oscillator 
strengths that are origin-independent also in finite basis-set calculations.  Note that the electric multipole transition moments
in the length and in the velocity representation are only equal in the basis set limit. However, the
calculation of higher-order transition moments requires sufficiently large basis sets anyway, so that the values in the 
length and in the velocity representation are usually in very good agreement.

For the calculation of X-ray absorption spectra in the following, we have employed the scheme of Stener \textit{et al.}\cite{stener_time_2003}
to allow only excitations from the relevant core orbital (see also Refs.~\cite{ray_description_2007,tsuchimochi_application_2008,
schmidt_assignment_2010,kovyrshin_state-selective_2010,liang_energy-specific_2011,lopata_linear-response_2012} for 
related schemes). 
For the Cl K-edge in TiCl$_4$, only excitations from the $1s$ orbital of one of the chlorine atoms were considered, while 
a frozen core was used for the other three chlorine atoms in order to obtain a localized core hole \cite{ray_description_2007}. 
For the Fe K-edge in vinylferrocene, only excitations from the iron $1s$ orbital were included.
All molecular structures were optimized using the BP86 exchange--correlation functional \cite{B88,Perdew86} 
and \textsc{Adf}'s TZP basis set. The TD-DFT calculations were performed using the BP86 functional and the 
TZ2P basis set and employed a fine numerical integration grid (integration accuracy 8). All calculations were performed 
with the scalar-relativistic zeroth-order regular approximation (ZORA) \cite{zora-1,zora-2,zora-3,zora-4}.

\section{Results and Discussion}
\label{sec:results}

\begin{figure}
\caption{Molecular structures of the model systems considered for the calculation of X-ray absorption spectra.
(a) Titaniumtetrachloride (TiCl$_{4}$) and (b) Vinylferrocene. The orientation of the molecules within the coordinate
system is also indicated.}
\label{fig:Mol}
\begin{center}
\includegraphics[width = 12cm]{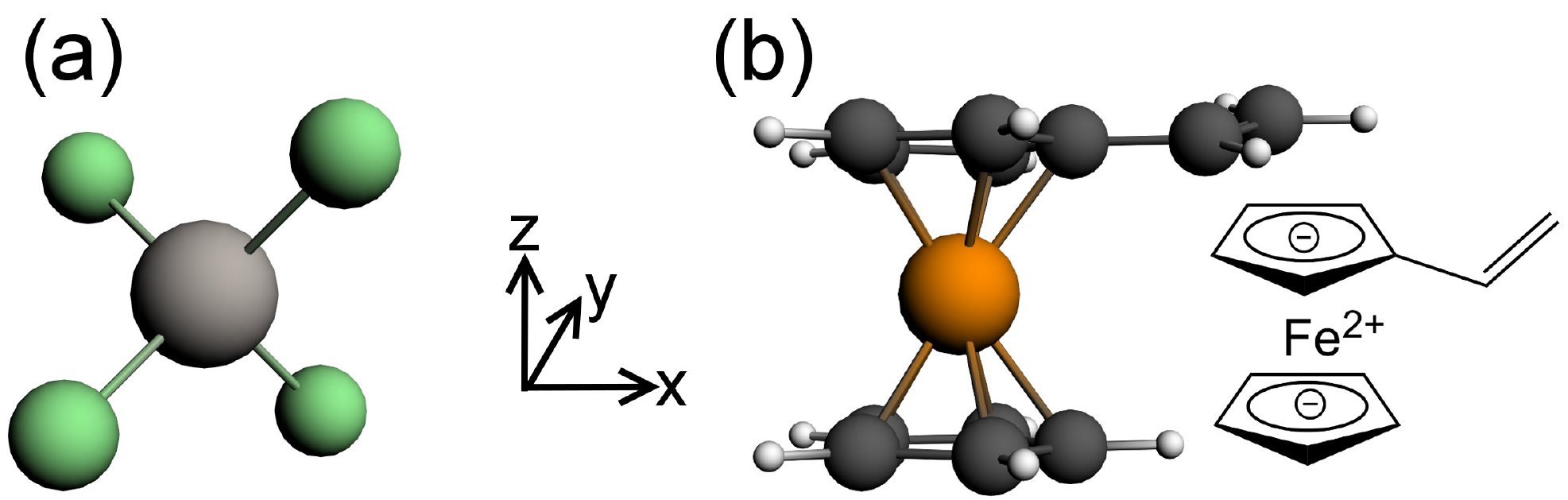}
\end{center}
\end{figure}

To illustrate the origin-independent calculation of quadrupole intensities in X-ray absorption spectroscopy (XAS) using the
theory derived above and to verify our implementation, we consider two test cases. As the first example, we use titaniumtetrachloride
TiCl$_4$ (see Fig.~\ref{fig:Mol}a for the molecular structure) and calculate the Cl K-edge XAS spectrum. This example was 
considered earlier in Refs.~\cite{stener_time_2003,debeer_george_metal_2005,debeer_george_time-dependent_2008}. 
For such ligand K-edge spectra, the prepeak transitions are dipole-allowed, and the second-order contribution to the 
oscillator strength should be small compared to the dipole contribution.

For the lowest-energy Cl K-edge excitation, the different contributions to the isotropically averaged oscillator strengths are 
calculated using Eqs~\eqref{eq:osc-avg-dipole}--\eqref{eq:osc-avg-mquad}, and are listed in Table~\ref{tab:Cl} for different 
choices of the origin. In addition, we included the oscillator strengths calculated using the approximation of 
Ref.~\cite{debeer_george_time-dependent_2008}, i.e.,
considering only the electric-dipole--electric-dipole, electric-quadrupole--electric-quadrupole, and the 
magnetic-dipole--magnetic-dipole contributions [cf.~Eq.~\eqref{eq:osc-neese}] as well as the full second-order 
oscillator strengths $f_{0n}^{(0)}+ f_{0n}^{(2)}$ [cf.~Eq.~\eqref{eq:osc-full}].

The most natural choice for the origin is the chlorine atom from which the $1s$-electron is excited. In this case,
the electric-dipole--electric-dipole contribution $f_{0n}^{(\mu^2)}$ to the oscillator strength is several orders of magnitude 
larger than all the second-order contributions, and the approximation of 
Ref.~\cite{debeer_george_time-dependent_2008} gives results that are identical to  
the full second-order oscillator strengths. In this example, the scheme suggested in 
Ref.~\cite{debeer_george_time-dependent_2008} to choose the origin such 
that the sum of the electric-quadrupole--electric-quadrupole $f_{0n}^{(Q^2)}$ and the magnetic-dipole--magnetic-dipole 
contributions $f_{0n}^{(m^2)}$ is minimized leads to an almost identical choice of the origin. Thus, this scheme is appropriate 
here. 

The situation changes if the origin is not placed at the chlorine atom. To demonstrate this, we moved the origin to the
titanium atom. Now, the electric-quadrupole--electric-quadrupole $f_{0n}^{(Q^2)}$ and the magnetic-dipole--magnetic-dipole 
contributions $f_{0n}^{(m^2)}$ increase significantly and become several times larger than the dipole oscillator strength 
$f_{0n}^{(\mu^2)}$. As a consequence, within the approximation of 
Ref.~\cite{debeer_george_time-dependent_2008} the oscillator strength increases by
more than a factor of two when shifting the origin from the chlorine to the titanium atom. However, also the magnitudes of
electric-dipole--electric-octupole and the electric-dipole--magnetic-quadrupole contributions, $f_{0n}^{(\mu O)}$ and
$f_{0n}^{(\mu\mathcal{M})}$, increase and since these have a negative sign, they exactly cancel the
increase of $f_{0n}^{(Q^2)}$ and $f_{0n}^{(m^2)}$. Thus, the full second-order oscillator strength remains unchanged.

In addition, we also shifted the origin away from the molecule by larger amounts. In particular, we used shifts of 
10~\AA, 50~\AA, and 100~{\AA} along the negative $x$-direction. Here, a similar observation can be made. The
electric-quadrupole--electric-quadrupole, $f_{0n}^{(Q^2)}$, and the magnetic-dipole--magnetic-dipole, $f_{0n}^{(m^2)}$,
contributions increase substantially, and for a shift of 100~\AA, the oscillator strength within the approximation of 
Ref.~\cite{debeer_george_time-dependent_2008}
is four orders of magnitude larger than for the origin at the chlorine atom. On the other hand, when including the
electric-dipole--electric-octupole and the electric-dipole--magnetic-quadrupole contributions, the full second-order
oscillator strengths are unchanged, even though the individual contributions differ.

\begin{table}
  \caption{X-ray absorption oscillator strength for TiCl$_{4}$, calculated for the lowest-energy transition at the 
  Cl K-edge (excitation energy  2755.6~eV). The total isotropically averaged oscillator strength and its different contributions are 
  given for different positions of the origin of the coordinate system.}
  \label{tab:Cl}
  \begin{center}
  {\small
  \begin{tabular}{cggggg}
    \hline\hline
origin 					&			&  \multicolumn{1}{c}{$\boldsymbol{O}$}	
											&  \multicolumn{1}{c}{$\boldsymbol{O} + \boldsymbol{a}$} 	
											&  \multicolumn{1}{c}{$\boldsymbol{O} + \boldsymbol{a}$}
											&  \multicolumn{1}{c}{$\boldsymbol{O} + \boldsymbol{a}$} \\
						& \multicolumn{1}{c}{Cl atom}	&  \multicolumn{1}{c}{Ti atom} 			
						& \multicolumn{1}{c}{$a_x=-10$ \AA} &  \multicolumn{1}{c}{$a_x=-50$ \AA}
						& \multicolumn{1}{c}{$a_x=-100$ \AA} \\
\hline
$\langle f_{0n}^{(\mu^2)} \rangle_{\text{iso}}$			
						& 4.57 \cdot 10^{-4}		& 4.57 \cdot 10^{-4}	& 4.58 \cdot 10^{-4}	& 4.57 \cdot 10^{-4}	& 4.58 \cdot 10^{-4} \\
$\langle f_{0n}^{(Q^2)} \rangle_{\text{iso}}$			
						& 6.04 \cdot 10^{-10}		& 7.24  \cdot 10^{-4}	& 1.09 \cdot 10^{-2}	& 3.65 \cdot 10^{-1}	& 1.35 \\
$\langle f_{0n}^{(m^2)} \rangle_{\text{iso}}$			
						& 6.61 \cdot 10^{-13}		& 1.21  \cdot 10^{-3}	& 1.80 \cdot 10^{-2}	& 3.58 \cdot 10^{-1}	& 2.23 \\
$\langle f_{0n}^{(\mu O)} \rangle_{\text{iso}}$			
						& -8.94 \cdot 10^{-7}		& -3.23 \cdot 10^{-4}	& -4.89 \cdot 10^{-3}	& -2.46 \cdot 10^{-1}  & -6.03 \cdot 10^{-1}	\\
$\langle f_{0n}^{(\mu\mathcal{M})} \rangle_{\text{iso}}$ 	
						& -9.88 \cdot 10^{-7}		& -1.61 \cdot 10^{-3}	& -2.40 \cdot 10^{-2}	& -4.77 \cdot 10^{-1}	 & -2.97 \\
\hline
$\langle f_{0n}^{(\mu^2)} 
  +f_{0n}^{(Q^2)} 
  +f_{0n}^{(m^2)} \rangle_{\text{iso}}$			
  						& 4.57 \cdot 10^{-4}		& 2.39 \cdot 10^{-3}	& 2.94 \cdot 10^{-2}	& 7.24  \cdot 10^{-1} & 3.58 \\
Full $\langle f_{0n}^{(0)} +  f_{0n}^{(2)} \rangle_{\text{iso}}$
						& 4.55 \cdot 10^{-4}		& 4.55 \cdot 10^{-4}	& 4.55 \cdot 10^{-4}	& 4.55 \cdot 10^{-4}  & 4.55 \cdot 10^{-4} \\
    \hline\hline 
  \end{tabular}}
  \end{center}
\end{table}

As a second example, we consider vinylferrocene, which is a ferrocene molecule bearing a vinyl substituent at one of the
cyclopentadienyl rings (see Fig.~\ref{fig:Mol}b for the molecular structure). Here, we consider the Fe K-edge XAS spectrum
and specifically the lowest-energy (prepeak) excitation, which is a $1s\rightarrow3d$ transition. In unsubstituted ferrocene, this 
prepeak excitation is dipole-forbidden for symmetry reasons, and its oscillator strength is solely due to the second-order 
contributions. In this case, the electric-quadrupole--electric-quadrupole and the magnetic-dipole-magnetic-dipole contributions 
become origin independent (see Section~\ref{sec:shift-equation}), whereas the remaining second-order contributions  
$f_{0n}^{(\mu O)}$ and $f_{0n}^{(\mu\mathcal{M})}$ vanish. However, in vinylferrocene this symmetry is lost and the 
lowest-energy transition gains a small dipole oscillator strength (for a detailed discussion, see Ref.~\cite{atkins_probing_2012}).

The oscillator strengths and their contributions calculated for the lowest-energy Fe K-edge excitation using different
choices of the origin are shown in Table~\ref{tab:Fe}. First, the most natural choice for the origin is the iron
atom. In this case, the electric-dipole--electric-dipole and the electric-quadrupole--electric-quadrupole contributions
to the oscillator strength are comparable in size. The remaining contributions are orders of magnitude
smaller. Therefore, the oscillator strength calculated with the approximation of 
Ref.~\cite{debeer_george_time-dependent_2008} is identical to the full second-order oscillator strength. 

To investigate the dependence on the origin, we shifted the origin far away from the molecule using a shift of
100~{\AA} in the negative $x$-direction, a shift of 100~{\AA} in the negative $z$-direction, and a shift of 50~{\AA} 
in both the negative $x$-direction and the negative $z$-direction. In all three cases, the electric-quadrupole--electric-quadrupole 
and the magnetic-dipole--magnetic-dipole contributions, $f_{0n}^{(Q^2)}$ and $f_{0n}^{(m^2)}$, increase by several orders of magnitude
compared to the calculation in which the origin is placed at the iron atom. As a result, the oscillator strengths
calculated with the approximation of Ref.~\cite{debeer_george_time-dependent_2008} also 
increase by up to five orders of magnitude. However, at the
same time the two remaining second-order contributions, i.e., the electric-dipole--electric-octupole contribution
$f_{0n}^{(\mu O)}$ and the electric-dipole--magnetic-quadrupole contribution $f_{0n}^{(\mu\mathcal{M})}$,
assume large negative values and exactly cancel the increase of $f_{0n}^{(Q^2)}$ and $f_{0n}^{(m^2)}$ such
that the total second-order oscillator strength remains origin independent.

Finally, we used the scheme suggested in Ref.~\cite{debeer_george_time-dependent_2008} for fixing 
the origin of the coordinate system, i.e.,
we chose the origin such that the sum of $f_{0n}^{(Q^2)}$ and $f_{0n}^{(m^2)}$ is minimized. In the situation
considered here, where the electric-dipole--electric-dipole and the electric-quadrupole--electric-quadrupole
contributions to the oscillator strengths are of similar size, this scheme moves the origin away from the iron
atom. The resulting shift is given in the caption of the last column of Table~\ref{tab:Fe}. As a consequence,
the oscillator strength within the approximation of Ref.~\cite{debeer_george_time-dependent_2008} decreases 
by ca.~30~\%. Again, this
decrease is compensated if the remaining second-order contributions are included. Thus, the scheme of
Ref.~\cite{debeer_george_time-dependent_2008} can lead to a spurious decrease of the oscillator strength 
in some cases. 
Previously, we found that this problem is even more severe in cases where the electric-dipole--electric-dipole
contribution to the oscillator strength is significantly smaller than the quadrupole oscillator strength \cite{atkins_probing_2012}.
However, if all second-order terms are included consistently the quadrupole oscillator strengths become origin-independent
and no special placement of the origin is necessary. 

\begin{table}
  \caption{X-ray absorption oscillator strength for Vinylferrocene, calculated for the lowest-energy transition at the 
  Fe K-edge (excitation energy  7051.3~eV). The total isotropically averaged oscillator strength and its different contributions are 
  given for different positions of the origin of the coordinate system.}
  \label{tab:Fe}
  \begin{center}
  {\small
  \begin{tabular}{cggggg}
    \hline\hline
origin 	& \multicolumn{1}{c}{$\boldsymbol{O}$}	& \multicolumn{1}{c}{$\boldsymbol{O} + \boldsymbol{a}$} 	
											& \multicolumn{1}{c}{$\boldsymbol{O} + \boldsymbol{a}$}
											& \multicolumn{1}{c}{$\boldsymbol{O} + \boldsymbol{a}$}
											& \multicolumn{1}{c}{$\boldsymbol{O} + \boldsymbol{a}$} \\
		& \multicolumn{1}{c}{Fe atom}	& \multicolumn{1}{c}{$a_x = -100$\AA} & \multicolumn{1}{c}{$a_z -100$ \AA} 	
								& \multicolumn{1}{c}{$a_x =-50$ \AA} 	& \multicolumn{1}{c}{$a_x = +0.128$ \AA} \\
		& 		& 				& & \multicolumn{1}{c}{$a_z =-50$ \AA}	& \multicolumn{1}{c}{$a_y = +0.195$ \AA} \\
		& 		& 				& 				& 				& \multicolumn{1}{c}{$a_z = +0.007$ \AA} \\
\hline
$\langle f_{0n}^{(\mu^2)}  \rangle_{\text{iso}}$			
		& 2.55 \cdot 10^{-6}	& 2.55 \cdot 10^{-6}	& 2.55 \cdot 10^{-6}	& 2.55 \cdot 10^{-6}	& 2.55 \cdot 10^{-6} \\
$\langle f_{0n}^{(Q^2)}  \rangle_{\text{iso}}$			
		& 3.09 \cdot 10^{-6}	& 8.63 \cdot 10^{-2}	& 7.30 \cdot 10^{-2}	& 4.47 \cdot 10^{-2}	& 1.28 \cdot 10^{-6} \\
$\langle f_{0n}^{(m^2)}  \rangle_{\text{iso}}$			
		& 1.14 \cdot 10^{-12}	& 2.29 \cdot 10^{-2}	& 9.04 \cdot 10^{-2}	& 5.87 \cdot 10^{-3}	& 1.56 \cdot 10^{-7} \\
$\langle f_{0n}^{(\mu O)}  \rangle_{\text{iso}}$			
		& -1.71 \cdot 10^{-8}	& -7.83 \cdot 10^{-2}	& -4.24 \cdot 10^{-2}	& -4.23 \cdot 10^{-2}	& 6.33 \cdot 10^{-7 } \\
$\langle f_{0n}^{(\mu\mathcal{M})}  \rangle_{\text{iso}}$ 	
		& -1.52 \cdot 10^{-8}	& -3.09 \cdot 10^{-2}	& -1.21 \cdot 10^{-1}	& -8.19 \cdot 10^{-3}	& 9.83 \cdot 10^{-7} \\
\hline
$\langle  f_{0n}^{(\mu^2)}
  +f_{0n}^{(Q^2)}
  +f_{0n}^{(m^2)}  \rangle_{\text{iso}}$			
  		& 5.64 \cdot 10^{-6}	& 1.09\cdot 10^{-1}	&1.63 \cdot 10^{-1} 	& 5.05 \cdot 10^{-2} 	& 3.99 \cdot 10^{-6} \\
Full $\langle  f_{0n}^{(0)}+ f_{0n}^{(2)}  \rangle_{\text{iso}}$	
		& 5.61 \cdot 10^{-6}	& 5.61 \cdot 10^{-6}	& 5.61 \cdot 10^{-6} 	& 5.61 \cdot 10^{-6} 	& 5.61 \cdot 10^{-6} \\
    \hline\hline 
  \end{tabular}}
  \end{center}
\end{table}

\section{Conclusions}
\label{sec:conclusion}

We have derived origin-independent expressions for calculating XAS intensities beyond the dipole
approximation. In particular, we have shown that for a consistent formulation, it is necessary to retain all
contributions to the oscillator strengths that are of the same order in the wave vector. This differs from the
previous approach \cite{debeer_george_time-dependent_2008}, in which the multipole expansion was
truncated for the transition moments. Here, two additional contributions to the second-order (quadrupole) 
oscillator strengths arise, which are cross-terms depending on products of electric-dipole and electric-octupole 
transition moments and of electric-dipole and magnetic-quadrupole transition moments, respectively.

Thus, the origin dependence of the sum of electric-quadrupole--electric-quadrupole and magnetic-dipole--magnetic-dipole 
contributions pointed out earlier \cite{debeer_george_time-dependent_2008} is not a fundamental limitation 
of the use of the multipole expansion. In fact, we could show 
that to arbitrary order in the wave vector, origin-independent expressions for the oscillator strengths are obtained if all
terms of the same order are included consistently. Consequently, within the multipole expansion it should always 
be possible to derive origin-independent expressions for physical observables. 

An origin-independent formalism for calculating quadrupole intensities is particularly important for studying ligand
and metal K-edge XAS spectra of transition metal complexes. To this end, we have implemented our theory for
calculating XAS spectra with TD-DFT, and applied it to two simple test cases.
Here, we want to stress that our results do not invalidate any previous results obtained with the formalism 
of Ref.~\cite{debeer_george_time-dependent_2008}. On the contrary, our test calculations showed that
the two additional contributions are negligible as long as the origin of the coordinate system is placed at
the atom where the core excitation occurs. However, with our origin-independent theory, it is no longer necessary 
to make sure that the origin is chosen appropriately. This is particularly important for cases where
the quadrupole intensity is larger than or comparable to the dipole contribution, where the scheme proposed 
in Ref.~\cite{debeer_george_time-dependent_2008} might place the origin far away from the relevant core orbital.
Moreover, it makes it possible to treat excitations from core orbitals that are delocalized over several atoms 
(e.g., for calculating ligand K-edge spectra or metal K-edge spectra in polynuclear transition metal complexes)
without the need to perform a transformation to localized core orbitals.

Of course, the theory presented here is not limited to TD-DFT, but can be employed for the calculation of
quadrupole intensities in combination with any quantum-chemical method capable of providing the required transition
moments. Moreover, it is not restricted to XAS spectroscopy, but is also applicable for calculating XES
intensities, for instance using the approach of Ref.~\cite{lee_probing_2010}.
Finally, we note that it becomes necessary to go beyond the dipole approximation, not only for short wavelengths, such as those 
employed in hard X-ray spectroscopy, but also for extended molecular systems. For describing the optical
response of an extended nanostructure in the visible spectrum, it becomes necessary to go beyond the dipole
approximation as well. Thus, the origin-independent formalism derived here will also be essential for predicting
optical properties of nanostructured materials, such as, for instance, metamaterials \cite{merlin_metamaterials_2009}.

\section*{Acknowledgments}

We thank Prof.~Wim Klopper (KIT) for inspiring this work and for helpful discussions.
Funding from the DFG-Center for Functional Nanostructures is gratefully acknowledged.

\begin{appendix}

\section*{Appendix}

\section{Length and velocity representation}
\label{app:comm}
 
\renewcommand{\theequation}{A-\arabic{equation}}
\setcounter{equation}{0}
 
To show how the electric-multipole moments in the velocity representation can be converted to those in
the conventional length representation, we use the following commutators of (products of) the Cartesian
components of the position operator with the molecular Hamiltonian given in Eq.~\eqref{eq:mol-hamiltonian},
\begin{align}
[r_{i,\alpha},\hat{H}_0] &= \frac{\text{i}\hbar}{m} \, \hat{p}_{i,\alpha} \,,\label{eq:comm1} \\[1ex]
[r_{i,\alpha} r_{i,\beta},\hat{H}_0] &= \frac{\text{i}\hbar}{m} (\hat{p}_{i,\alpha} r_{i,\beta} + r_{i,\alpha}\hat{p}_{i,\beta})\,, \label{eq:comm2} \\[1ex]
[r_{i,\alpha} r_{i,\beta} r_{i,\gamma}, \hat{H}_0] 
  &= \frac{\text{i}\hbar}{m} (\hat{p}_{i,\alpha} r_{i,\beta} r_{i,\gamma} + r_{i,\alpha} \hat{p}_{i,\beta} r_{i,\gamma} 
        + r_{i,\alpha} r_{i,\beta} \hat{p}_{i,\gamma})\,. \label{eq:comm3}
\end{align}

Next, we employ that the matrix elements of the commutator of an operator $\hat{A}$ and $\hat{H}_0$ are 
given by
\begin{equation}
  \trel{[\hat{A},\hat{H}_0]}
  =  \trel{\hat{A}\hat{H}_0 - \hat{H}_0\hat{A}}
  = E_n \trel{\hat{A}} - E_0 \trel{\hat{A}} = E_{0n} \trel{\hat{A}}
\end{equation}
Here, it is important to point out that this relation is only valid for the exact eigenfunctions
of $\hat{H}_0$ and that it only holds approximately for approximate wavefunctions.

Now, we can use these results to obtain
\begin{align}
  \trel{\hat{p}_{i,\alpha}}
  =  \frac{m}{\text{i}\hbar} \trel{[r_{i,\alpha},\hat{H}_0]}
  =  - \text{i} E_{0n} \frac{m}{\hbar} \trel{r_{i,\alpha}}  
\end{align}
and get for the electric-dipole transition moments
\begin{align}
  \trel{\hat{\mu}_\alpha^p} 
  = \sum_i \frac{e}{m} \trel{\hat{p}_{i,\alpha}}
  =  - \text{i} \frac{E_{0n}}{\hbar} e \sum_i \trel{r_{i,\alpha}}
  =  - \text{i} \frac{E_{0n}}{\hbar}   \trel{\hat{\mu}_\alpha}.
\end{align}
Similarly, we find for the electric-quadrupole transition moments
\begin{align}
  \trel{\hat{Q}_{\alpha\beta}^p} 
  =  - \text{i} \frac{E_{0n}}{\hbar} \trel{\hat{Q}_{\alpha\beta}} 
\end{align}
and for the electric-octupole transition moments
\begin{align}
  \trel{\hat{O}_{\alpha\beta\gamma}^p} 
  =  - \text{i} \frac{E_{0n}}{\hbar} \trel{\hat{O}_{\alpha\beta\gamma}}.
\end{align}

\section{Antisymmetric second-order term}
\label{app:antisymm-2nd-order}

\renewcommand{\theequation}{B-\arabic{equation}}
\setcounter{equation}{0}

First, the matrix elements in the antisymmetric term in Eq.~\eqref{eq:second-order-matrixelements}, can be split as
\begin{align}                                                              
  &2(\boldsymbol{k}\cdot\boldsymbol{r}_i)(\boldsymbol{k}\cdot\boldsymbol{r}_i)(\hat{\boldsymbol{p}}_i\cdot\be)
                    - (\boldsymbol{k}\cdot\boldsymbol{r}_i)(\boldsymbol{k}\cdot\hat{\boldsymbol{p}}_i)(\boldsymbol{r}_i\cdot\be)
                    - (\boldsymbol{k}\cdot\hat{\boldsymbol{p}}_i)(\boldsymbol{k}\cdot\boldsymbol{r}_i)(\boldsymbol{r}_i\cdot\be) \nonumber \\
  =&\ (\boldsymbol{k}\cdot\boldsymbol{r}_i)(\boldsymbol{k}\cdot\boldsymbol{r}_i)(\hat{\boldsymbol{p}}_i\cdot\be)
                    - (\boldsymbol{k}\cdot\boldsymbol{r}_i)(\boldsymbol{k}\cdot\hat{\boldsymbol{p}}_i)(\boldsymbol{r}_i\cdot\be)  \nonumber \\
     &+ (\boldsymbol{k}\cdot\boldsymbol{r}_i)(\boldsymbol{k}\cdot\boldsymbol{r}_i)(\hat{\boldsymbol{p}}_i\cdot\be)    
                    - (\boldsymbol{k}\cdot\hat{\boldsymbol{p}}_i)(\boldsymbol{k}\cdot\boldsymbol{r}_i)(\boldsymbol{r}_i\cdot\be).                                               
\end{align}
For both terms, we can employ that both $(\boldsymbol{k}\cdot\boldsymbol{r}_i)$ and $(\hat{\boldsymbol{p}}_i \cdot \be)$ 
commute because $\boldsymbol{k}$ and $\be$ are orthogonal, and subsequently use the vector identity of Eq.~\eqref{eq:vectid}.
In the same fashion as for the antisymmetric first-order contribution, we then obtain for the first term,
\begin{align}
  &(\boldsymbol{k}\cdot\boldsymbol{r}_i) \big[ (\boldsymbol{k}\cdot\boldsymbol{r}_i)(\hat{\boldsymbol{p}}_i\cdot\be) 
      - (\boldsymbol{k}\cdot\hat{\boldsymbol{p}}_i)(\boldsymbol{r}_i\cdot\be) \big]
  \nonumber \\
  &\qquad= (\boldsymbol{k}\cdot\boldsymbol{r}_i)(\boldsymbol{k} \times \be) \cdot (\boldsymbol{r}_i \times \hat{\boldsymbol{p}}_i) 
                 = (\boldsymbol{k} \times \be) \cdot (\boldsymbol{k}\cdot\boldsymbol{r}_i) \,(\boldsymbol{r}_i \times \hat{\boldsymbol{p}}_i),
\end{align}
and for the second term, we get,
\begin{align}
  &(\boldsymbol{k}\cdot\boldsymbol{r}_i)(\boldsymbol{k}\cdot\boldsymbol{r}_i)(\hat{\boldsymbol{p}}_i\cdot\be) 
       - (\boldsymbol{k}\cdot\hat{\boldsymbol{p}}_i)(\boldsymbol{k}\cdot\boldsymbol{r}_i)(\boldsymbol{r}_i\cdot\be)
  \nonumber \\
  &\qquad = (\boldsymbol{k}\cdot\boldsymbol{r}_i)(\hat{\boldsymbol{p}}_i\cdot\be)(\boldsymbol{k}\cdot\boldsymbol{r}_i)
       - (\boldsymbol{k}\cdot\hat{\boldsymbol{p}}_i)(\boldsymbol{r}_i\cdot\be)(\boldsymbol{k}\cdot\boldsymbol{r}_i)
  \nonumber \\
  &\qquad = \big[(\boldsymbol{k}\cdot\boldsymbol{r}_i)(\hat{\boldsymbol{p}}_i\cdot\be)
       - (\boldsymbol{k}\cdot\hat{\boldsymbol{p}}_i)(\boldsymbol{r}_i\cdot\be)\big](\boldsymbol{k}\cdot\boldsymbol{r}_i)
  \nonumber \\
  &\qquad = (\boldsymbol{k} \times \be) \cdot (\boldsymbol{r}_i \times \hat{\boldsymbol{p}}_i)(\boldsymbol{k}\cdot\boldsymbol{r}_i).
\end{align}
Altogether, we arrive at
\begin{align}
  &2(\boldsymbol{k}\cdot\boldsymbol{r}_i)(\boldsymbol{k}\cdot\boldsymbol{r}_i)(\hat{\boldsymbol{p}}_i\cdot\be)
                    - (\boldsymbol{k}\cdot\boldsymbol{r}_i)(\boldsymbol{k}\cdot\hat{\boldsymbol{p}}_i)(\boldsymbol{r}_i\cdot\be)
                    - (\boldsymbol{k}\cdot\hat{\boldsymbol{p}}_i)(\boldsymbol{k}\cdot\boldsymbol{r}_i)(\boldsymbol{r}_i\cdot\be) \nonumber \\
 =&\ (\boldsymbol{k} \times \be) \cdot \big[ (\boldsymbol{k}\cdot\boldsymbol{r}_i) \cdot (\boldsymbol{r}_i \times \hat{\boldsymbol{p}}_i) 
 + (\boldsymbol{r}_i \times \hat{\boldsymbol{p}}_i)(\boldsymbol{k}\cdot\boldsymbol{r}_i) \big].
\end{align}

\section{Origin independence in arbitrary order}
\label{app:origin-higher-order}

\renewcommand{\theequation}{C-\arabic{equation}}
\setcounter{equation}{0}

In Section~\ref{sec:osc-origin-indep}, we showed explicitly that the second-order oscillator strengths are  independent of
the choice of the origin. Here, we prove that this still holds for an arbitrary order. From the definition of the full transition
moments [Eq.~\eqref{eq:t-full}], we find for its change upon shifting the origin from $\boldsymbol{O}$ to $\boldsymbol{O}
+\boldsymbol{a}$,
\begin{align}
  T(\boldsymbol{O}+\boldsymbol{a}) 
  = \exp({\rm i}\boldsymbol{k} \cdot \boldsymbol{a}) T(\boldsymbol{O}) 
  = \Big( \sum_{n=0}^\infty \frac{{\rm i}^n}{n!} (\boldsymbol{k} \cdot \boldsymbol{a})^n \Big) \Big( \sum_{n=0}^\infty T^{(n)}(\boldsymbol{O}) \Big),
\end{align}
and can identify the terms that are of order $m$ in the wave vector,
\begin{align}
  T^{(m)}(\boldsymbol{O}+\boldsymbol{a}) 
  = \sum_{n=0}^m \frac{{\rm i}^n}{n!} (\boldsymbol{k} \cdot \boldsymbol{a})^n \ T^{(m-n)}(\boldsymbol{O}).
\end{align}
For the oscillator strengths, the terms that are of order $m$ in the wave vector are,
\begin{align}
  f^{(m)}(\boldsymbol{O}) = \frac{2m_e}{e^2E_{0n}} \, \sum_{n=0}^m  T^{(n)}(\boldsymbol{O}) \big[T^{(m-n)}(\boldsymbol{O})\big]^*,
\end{align}
and when the origin of the coordinate system is shifted, this becomes
\begin{align}
  f^{(m)}(\boldsymbol{O} + \boldsymbol{a}) 
  &= \frac{2m_e}{e^2E_{0n}} \,
        \sum_{n=0}^m  T^{(n)}(\boldsymbol{O}  + \boldsymbol{a}) \big[T^{(m-n)}(\boldsymbol{O} + \boldsymbol{a})\big]^* \nonumber \\
  &= \frac{2m_e}{e^2E_{0n}} \,
        \sum_{n=0}^m \Bigg[ \sum_{p=0}^n \frac{{\rm i}^p}{p!}(\boldsymbol{k} \cdot \boldsymbol{a})^p \ T^{(n-p)}(\boldsymbol{O}) \Bigg] 
                                  \Bigg[\sum_{q=0}^{m-n} \frac{(-{\rm i})^q}{q!} (\boldsymbol{k} \cdot \boldsymbol{a})^q \ \big[T^{(m-n-q)}(\boldsymbol{O}) 
                                  \Big]^*\Bigg] \nonumber \\
  &= \frac{2m_e}{e^2E_{0n}} \,
        \sum_{n=0}^m \ \sum_{p=0}^n\sum_{q=0}^{m-n} \ (-1)^q \frac{{\rm i}^{p+q}}{p!q!} \   
                                   (\boldsymbol{k} \cdot \boldsymbol{a})^{p+q} \ T^{(n-p)}(\boldsymbol{O}) \, \big[T^{(m-n-q)}(\boldsymbol{O}) \Big]^*.
\end{align}
Now we eliminate $p$ and $q$ by introducing the new indices $r=p+q$ and $s=n-p$ to arrive at
\begin{align}
  \label{eq:sum1}
  f^{(m)}(\boldsymbol{O} + \boldsymbol{a}) 
  =& \frac{2m_e}{e^2E_{0n}} \, \sum_{n=0}^m \ \sum_{s=0}^n\sum_{r=n-s}^{m-s} \ \frac{(-1)^{r-n+s}  \, {\rm i}^r}{(n-s)!(r-n+s)!} \   
                                   (\boldsymbol{k} \cdot \boldsymbol{a})^{r} \ T^{(s)}(\boldsymbol{O}) \, \big[T^{(m-r-s)}(\boldsymbol{O}) \Big]^* \nonumber \\
\end{align}
The three sums can be rewritten and put in a different order, which leads to
%
\begin{align}
  f^{(m)}(\boldsymbol{O} + \boldsymbol{a}) 
  =& \ \frac{2m_e}{e^2E_{0n}} \, \sum_{r=0}^m (\boldsymbol{k} \cdot \boldsymbol{a})^{r} \ 
        \sum_{s=0}^{m-r} T^{(s)}(\boldsymbol{O}) \, \big[T^{(m-r-s)}(\boldsymbol{O}) \Big]^* \
        \sum_{n=s}^{s+r} \frac{(-1)^{r-n+s}  \ {\rm i}^r}{(n-s)!(r-n+s)!}   
                                    \nonumber \\
 =& \ f^{(m)}(\boldsymbol{O})  \nonumber \\
  &+ \frac{2m_e}{e^2E_{0n}} \, \sum_{r=1}^m (\boldsymbol{k} \cdot \boldsymbol{a})^{r} \ 
        \sum_{s=0}^{m-r} T^{(s)}(\boldsymbol{O}) \, \big[T^{(m-r-s)}(\boldsymbol{O}) \Big]^* \
         \sum_{t=0}^{r}  \frac{(-1)^{r-t} \ {\rm i}^r}{t!(r-t)!} 
\end{align}
That the summation here is equivalent to the one in Eq.~\eqref{eq:sum1} can be seen easily by considering
the six inequalities corresponding to the sums in the two cases and showing that these are equivalent.  In
the second line above, we have taken the term  $m=0$ out for the first sum and introduced the new index 
$t=n-s$.

Finally, we can use the binomial theorem to realize that,
\begin{equation}
  0 = ({\rm i} - {\rm i})^r = \sum_{t=0}^r \frac{r!}{t!(r-t)!} \, {\rm i}^{t} (-{\rm i})^{r-t}
  = r! \sum_{t=0}^r \frac{(-1)^{r-t} \  {\rm i}^r }{t!(r-t)!}, 
\end{equation}
that is, the last term in the above equation is zero and, thus, we have shown that in any order $k$, the
oscillator strengths are origin-independent.

%
%

\end{appendix}

\end{document}